# Dorsal lateral geniculate substructure in the Long-Evans rat: A cholera toxin B-subunit study

**Claire B. Discenza[1] and Pamela Reinagel[2,*]**
1. Department of Neuroscience, School of Medicine, University of California, San Diego, La Jolla, CA, USA.
2. Section of Neurobiology, Division of Biological Sciences, University of California, San Diego, La Jolla, CA, USA.
*to whom correspondence should be addressed:

## ABSTRACT

This study describes the substructure of the dorsal lateral geniculate nucleus of the thalamus of the pigmented rat (*Rattus norvegicus*) based on the eye-of-origin of its retinal ganglion cell inputs. We made monocular intra-ocular injections of the B-subunit of cholera toxin (CTB), a sensitive anterograde tracer, in three adult male Long-Evans rats. In four additional subjects, we injected fluorophor-conjugated CTB in both eyes, using a different fluorophor in each eye. Brains of these subjects were fixed and sectioned, and the labeled retinal ganglion cell termini were imaged with wide-field sub-micron resolution slide scanners. Retinal termination zones were traced to reconstruct a three dimensional model of the ipsilateral and contralateral retinal termination zones in the dLGN on both sides of the brain. The dLGN volume was 1.58 ±0.094 $mm^3$, comprising 70 ± 3% the volume of the entire retinorecipient LGN. We find the retinal terminals to be well-segregated by eye of origin. We consistently found three or four spatially separated ipsilateral-recipient zones within each dLGN, rather than the single compact zone expected. It remains to be determined whether these subdomains represent distinct functional sublaminae.

# INTRODUCTION

In mammals, the retinal ganglion cells (RGC) that contribute to cortical vision send their projections to the dorsal lateral geniculate nucleus of the thalamus (dLGN). Ipsilateral and contralateral projections remain segregated at the level of the dLGN (Matteau et al., 1993). In many species, the projections of distinct ganglion cell types further segregate into discrete dLGN laminae, each containing a retinotopic map of visual space (Bishop et al, 1962; Laties and Sprague, 1966; Garey and Powell, 1968; Kinston et al., 1969; Sanderson et al., 1971a,b; Jones, 2007, for review). Although species differ in their retinal ganglion cell classes and the in the details of their dLGN lamination, the segregation of information into anatomically separate parallel processing streams is a conserved organizing principle of the dLGN (Roy et al., 2009; Cleland et al., 1971; So, 1990; Kaas et al., 1972; Sherman and Guillery 2001 for review).

For example, the dLGN of the macaque monkey (*Macaca mulatta*) contains six layers, each receiving inputs from a different subset of retinal ganglion cells (parasol or midget; ON or OFF; ipsilateral or contralateral) (Schiller and Malpeli, 1978; Shapley and Perry 1986; Szmajda et al., 2006; Connolly and Van Essen,1994; Malpeli and Baker, 1975; Murray et al., 2008). In the cat (*Felis catus)*, six layers have been distinguished in the dLGN, receiving inputs from different subsets of RGC types (X, Y or W) and segregated by eye of origin (Guillery 1970; Sherman and Spear, 1982; Shapley and Perry 1986). The ferret (*Mustela putorius furo*) is similar to cat with further sub-lamination of ON and OFF types (Stryker & Zahs 1983). In the California ground squirrel (*Spermophilus beecheyi*), a diurnal rodent, the retinal projections form three layers with alternating eye of origin in the dLGN; six sublaminae have been distinguished (Roe et al., 1989).

It remains unclear, however, whether the dLGN is as highly organized in nocturnal rodents such as the rat and mouse, which lack obvious lamination in the Nissl preparation (see Jones, 2007, for review). Nevertheless, these nuclei are not homogenous. In the mouse (*Mus musculus*), distinct functional classes of RGCs have been found to project to distinct layers in the dLGN (Huberman et al.,2008; Huberman et al., 2009). In the rat (*Rattus norvegicus*), the nucleus has been subdivided into two general regions by anatomy and physiology: an outer lateral 'shell' and an inner medial 'core' (see Reese,1988 for review). These two regions differ in that they receive projections from differing populations of morphological ganglion cell types (Fukuda, 1977; Hickey et al., 1976; Bunt et al., 1974; Brauer et al.,1979) and contain distinct morphological classes of relay cells and termini (Bartlett and Smith, 1999; Lund and Cunningham,1972). In addition, the outer 'shell' receives input from the optic tectum (Reese, 1984), and the inner 'core' contains a smaller internal region which receives termini emanating from the ipsilateral eye (Reese and Cowey, 1983). It has been shown that these segregated zones contain their own retinotopic map of visual space, although only the contralateral outer 'shell' region is known to contain a complete map (Reese and Jeffrey, 1983; Reese, 1988; Montero et al.,1968). Most studies report segregation of inputs by eye of origin in the dLGN of pigmented rats (Reese, 1988; Guido, 2006; but see Hayhow et al., 1962; Greive, 2005).

In the current study, we used Cholera Toxin B subunit, an efficient anterograde RGC tracer (Angelucci et al. 1996; Matteau et al., 2003; Reiner et al., 1996), to visualize retinal termination zones in the rat dLGN. First we measured the volume of the dLGN relative to other retinorecipient structures: the optic tectum and the ventral lateral geniculate (vLGN). Second, we reconstructed a three dimensional model of the ipsilateral and contralateral projections to the dLGN to determine if there is more than one topologically discrete projection zone for either eye. Third, we determined the accuracy of segregation into eye-specific domains within the dLGN.

# MATERIALS AND METHODS

**Subjects**

Seven normal adult male Long Evans rats (Harlan Laboratories, Inc.) were used in this study. Rats 1-4 received binocular injections of fluorescently labeled Cholera Toxin B subunit (CTB, see below) at 3-4 months of age (370-440 grams). Subjects 5-7 received monocular injections of CTB at 6-7 months of age (490-670 grams). All subjects were maintained on a 12-hr light/dark cycle with free access to food and water. All procedures were supervised and approved by the Institutional Care and Use Committee at the University of California, San Diego.

**Intra-ocular Injections**

The B subunit of the Cholera Toxin complex (CTB) has been shown to be a highly-sensitive anterograde tracer for RGCs (Matteu et al., 2003; Rainer et al., 1996; Angelucci et al., 1996), and therefore the preferred tracer for this study. Rats were first anesthetized with 2-5% isoflurane mixed with oxygen at a flow rate of one liter per min., using an isoflurane vaporizer (Smiths Medical, Dublin, OH.) While maintained at the appropriate level of anesthesia, subjects were subcutaneously injected with buprenorphine (0.06 mg/kg rat weight). Subjects then received, via syringe, 5-6 2µl injections of either unconjugated CTB in one eye, or fluorophor-conjugated CTB in both eyes.

The monocular injections were administered into the vitreous chamber of the left eye only, and comprised a 1% CTB solution (List Biological Laboratories, Inc., Campbell, CA) mixed with 2% dimethyl sulfoxide (DMSO) diluted in sterile water. In the fluorescent case, rats were injected with two different fluorophor CTB conjugations, one in the vitreous chamber of each eye. A 1mg/ml dilution of Alexa Fluor 488-conjugated CTB (Molecular Probes Inc., Eugene, OR) in PBS was injected into the left eye, and a similar dilution of Alexa Fluor 594-conjugated CTB was injected into the right eye (Molecular Probes Inc., Eugene, OR).

We waited 5-7 days post-injection before perfusion to allow for transport of tracer to the retinal termini (Wu et al., 1999). During this post-injection survival period subjects received twice-daily buprenorpnine injections for a minimum of 2-3 days, continuing as needed until sacrificed for perfusion and histology.

**Perfusion and Histology**

Five to seven days post-injection, all rats were euthanized with an overdose of isofluorane and perfused transcardially with 0.1M phosphate-buffered saline (PBS; pH 7.4) followed by 4% paraformaldehyde in PBS. After removal, brains were further fixed in 4% paraformaldehyde for at least three days, after which they were then soaked in a 30% sucrose PBS buffer solution for cryoprotection prior to slicing. Brains were sliced on a freezing microtome (Microm International GmbH, Waldorf, Germany); brains from non-conjugated monocularly-injected rats were sliced at 30µm in one of the three planes, and binocularly-labeled brains were sliced at 25µm coronally.

Fluorescent samples were sliced, separated into four series and mounted with Prolong Gold anti-fade reagent medium (Molecular Probes Inc., Eugene, OR) on charged slides (Thermo Fisher Scientific Inc., Pittsburgh, PA) and covered with a cover slip. After the initial round of imaging, slides were soaked to remove the cover slip, photo-bleached, and stained with NeuroTrace 500/525 nm green fluorescent Nissl stain (Molecular Probes Inc., Eugene, OR) for other analyses described elsewhere (Discenza, 2011).

Non-fluorescent tissue samples were processed according to the method described by Angelucci et al. (1996) and Matteo et al. (2003). In summary, tissue was rinsed in phosphate buffered saline, and then incubated and rotated at 4° Celsius overnight in a primary antibody solution of 0.1% Triton X-100, 5% normal rabbit serum, and a 1:1000- 1:2000 dilution of biotinylated goat anti-rabbit CTB (List Biological Laboratories, Inc., Campbell, CA, cat #103B) in phosphate buffered saline. After rinsing again with phosphate buffered saline, tissue was then incubated and slowly rotated for one hour at room temperature in the secondary antibody solution consisting of a 1:1000 dilution of Vectastain biotinylated IgG (Vector Labs, Burlingame, CA #PK-4005) with 0.3% Triton X-100 in phosphate buffered saline. Finally, after a third set of rinses, tissue was incubated in a tertiary antibody solution made using the Vectastain ABC kit ElitePK-6100 kit (Vector Labs, Burlingame, CA). Tissue was incubated in a complexed avidin-biotin-peroxidase solution diluted to 1:1000 in phosphate buffered saline with 0.3% Triton X-100 and additional 2% NaCl. To visualize the CTB, the tissue was rinsed in buffer and soaked in a 1:3000 hydrogen peroxide phosphate buffer solution with 0.125 mg/ml Diaminobenzidine (DAB) for approximately 1 minute, or until cells reacted. Tissue was rinsed, mounted on gel-coated slides (Thermo Fisher Scientific Inc., Pittsburgh, PA), enhanced with 4% osmium, and coverslipped. One series from each brain was reacted with DAB alone, one series was counterstained with Geimsa as well as DAB, and another series was Nissl-stained for other analyses described elsewhere (Discenza, 2011).

**Imaging**

Fluorescent samples were imaged on the NanoZoomer 2.0 HT digital slide scanner (Hamamatsu Photonics, Japan). Slides were imaged at 20x resolution (0.46 $\mu m^2$/pixel) using the fluorescent cube (DAPI/Fluorescein isothiocyanate/TexasRed). The non-fluorescent DAB/Geimsa series were scanned using Aperio Scanscope XT digital slide scanner (Aperio Technologies Inc., Vista, CA; Burnham Institute, La Jolla, CA) at 20x resolution (0.5 $\mu m^2$/pixel), aligned using ImageJ software (Abramoff et al., 2004), and analyzed using custom software written in MATLAB (2008a-2010a, The MathWorks, Natick, MA). We confirmed that CTB filling of ganglion cells was complete by verifying uniform and complete staining in the optic tectum and across the dLGN.

**Tracing and 3D Reconstructions**

The dLGN termination zones were hand-traced over the high-resolution digitized images of filled RGC termini, using the software Neurolucida (MBF Biosciences, Inc., Williston, VT). The Rat Brain Stereotaxic Atlas (Paxinos and Watson, 1998, fourth edition) was used for initial identification of nuclei.

Outlines were traced delineating: a) the entire dLGN, tracing the outer edge of both ipsilateral and contralateral termini, b) ipsilateral subregions, or contiguous regions containing puncta from the ipsilateral RGC termini, c) and the 'holes' in the contralateral zones, or contiguous regions within the dLGN lacking contralateral retinal projections. Termination zones were traced while visualizing one fluorophor at a time. The aim was to encircle contiguous regions of retinal projections, and to separate these regions only when the distance between them was large compared to the termini density within the regions.

Projection zones were outlined manually according to defined tracing criteria (Figure 1). First, fibers of passage were observed but not included when defining the outline of a region. Fibers of passage typically fluoresced more faintly, and formed extended axonal shapes and not dense bright groups of puncta as did termini (Figure 1, **A**). Secondly, areas of very low density were ignored, for example, areas containing fewer than one or two termini in 300 $\mu m^2$ (Figure 1, **B**). Areas of high density, where puncta were either overlapping or up to 10 μm apart (Figure 1, **C**), were traced, as were areas of low density, where puncta were10-20 μm apart (Figure 1, **D**). Outlines were drawn approximately 5 μm around the "outer boundary" of a termination zone, here defined as the outer termini of a zone where the next nearest neighboring puncta or zone is approximately 20 μm away.

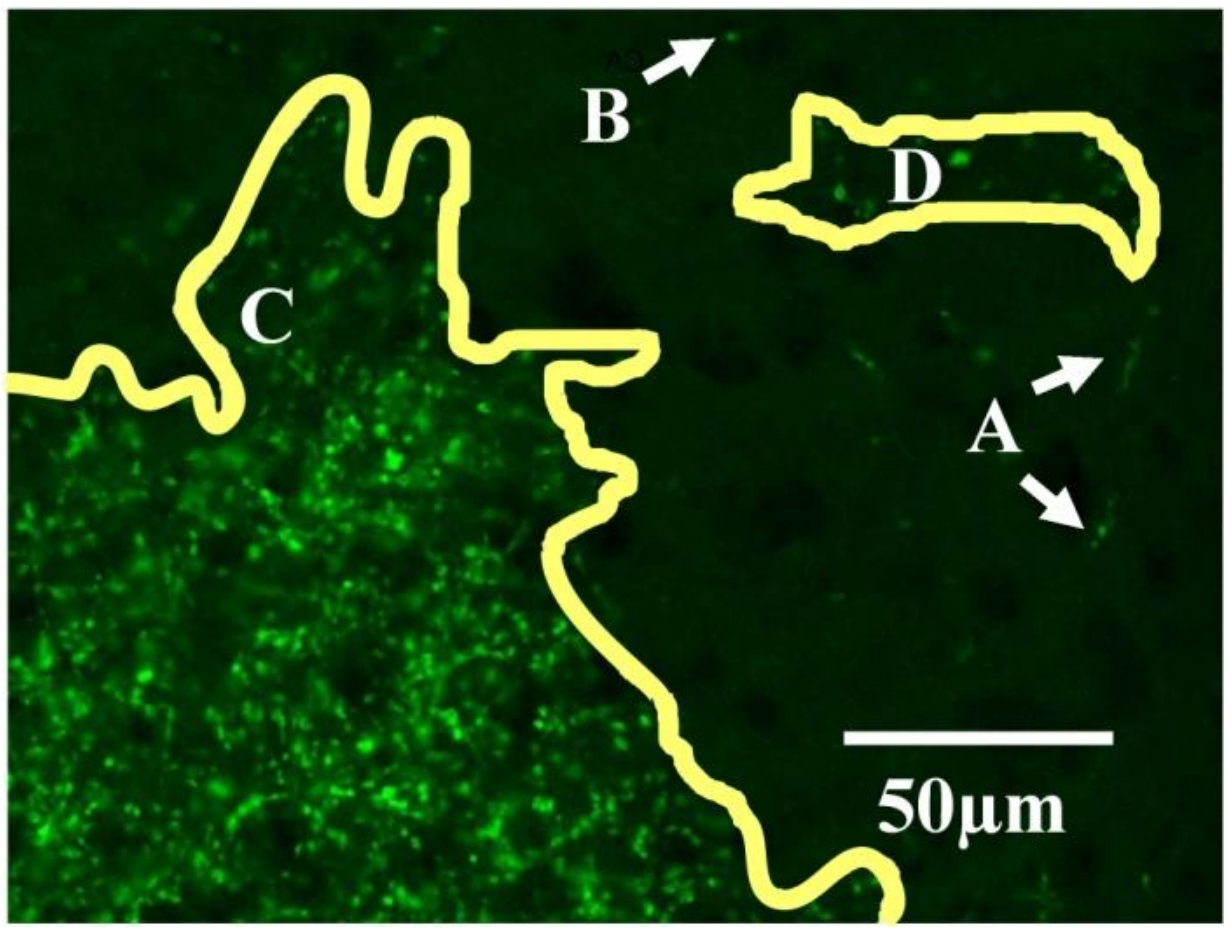

**Figure 1. Tracing criteria for termination zones.** The boundaries of retinal projection zones were manually traced from wide field (whole brain), high resolution (0.46μm/pixel) digitally scanned images, such as this field of view showing ipsilateral projections in part of the dLGN. At this resolution we can distinguish fibers of passage (**A**) from retinal termini; fibers of passage were ignored when outlining projection zones. Isolated single termini (**B**) were occasionally observed, and were ignored when outlining projection zones. Outlines were drawn around clusters of termini by smoothly connecting the outermost retinal termini within the cluster (**C**, **D**). In the more densely-populated projection zone shown here (**C**), termini were overlapping to 10 μm apart. Clusters of termini that were far apart relative to the inter-terminal distances within the clusters were assigned to separate outlines (**C** vs. **D**). Separate outlines could potentially be assigned to the same subregion in the 3-dimensional reconstruction (see Figure 2).

We traced all sections throughout one specimen (Rat 1) and every fourth section throughout the remaining three fluorescent specimens (Rats 2-4). The outlines were aligned with Neurolucida, using ventricles, blood vessels, fiduciary pin marks, and dLGN outlines as landmarks. The aligned outlines were then stacked in Neurolucida to create a three dimensional model of the dLGN. The same procedures and criteria were used to trace projection zones in the non-fluorescent, DAB-stained specimens from monocularly injected rats (Rats 5-7), but using custom Matlab software. Results did not differ in the two data sets.

**Assignment of termini to subdomains**

Using the three-dimensional models, which can be rotated and examined from all angles, individual subdomains of retinal termination were distinguished by defined criteria in the four binocularly labeled specimens (Rats 1-4). Previous reports described only one compact ipsilateral termination zone in the rat dLGN, so our null hypothesis was that all ipsilateral projection outlines on each section through the dLGN were part of the same ipsilateral subdomain in three dimensions, despite appearing separated in the two dimensional plane. Therefore we were conservative in postulating separate subdomains, requiring that overlapping outlines were found in adjacent sections spanning at least 200 μm, and separation of at least 75 μm between it and its closest neighboring subdomain. Thus for specimen 1, in which every 25 μm section was traced, we required 8 consecutive sections to contain an overlapping outlined termination zone, and at least 3 sections distance from the closest neighboring subdomain. For specimens 2-4, every fourth 25 μm section was traced, and we required 3 in a row to contain an overlapping region.

The overlap criteria for grouping outlines depended on whether the termination zones were compact or sparse. When dLGN regions contained dense termini (overlapping or up to 10 μm apart, Figure 1, **C**) we required that each consecutive stacked outline overlap with its neighboring section by at least 10% in area (Figure 2, **A**). When an outlined region contained sparse termini (Figure 1, **D**), its boundary was by nature ill-defined due to low sampling. For this reason, the distance to the closest terminal in the next section had to be large relative to the within section distance for the sections to be considered a separate subdomain. Therefore, while a sparsely-populated subdomain still had to contain at least 8 sections and be separated by approximately 75 μm from its closest neighbor, the alignment criteria was relaxed, and the outlines only had to overlap 1% with adjacent sections (Figure 2, **B**).

In some cases, outlines aligned less than 1% with neighboring sections, or did not fall in groups of 8 or more outlines. When a loose stack of these “stray outlines” was more closely grouped together than with any neighboring subdomain, these outlines were assigned to a “diffuse subdomain” (see Figure 2, **C**). Otherwise, these stray outlines were incorporated into the closest neighboring subdomain (Figure 2, **D**).

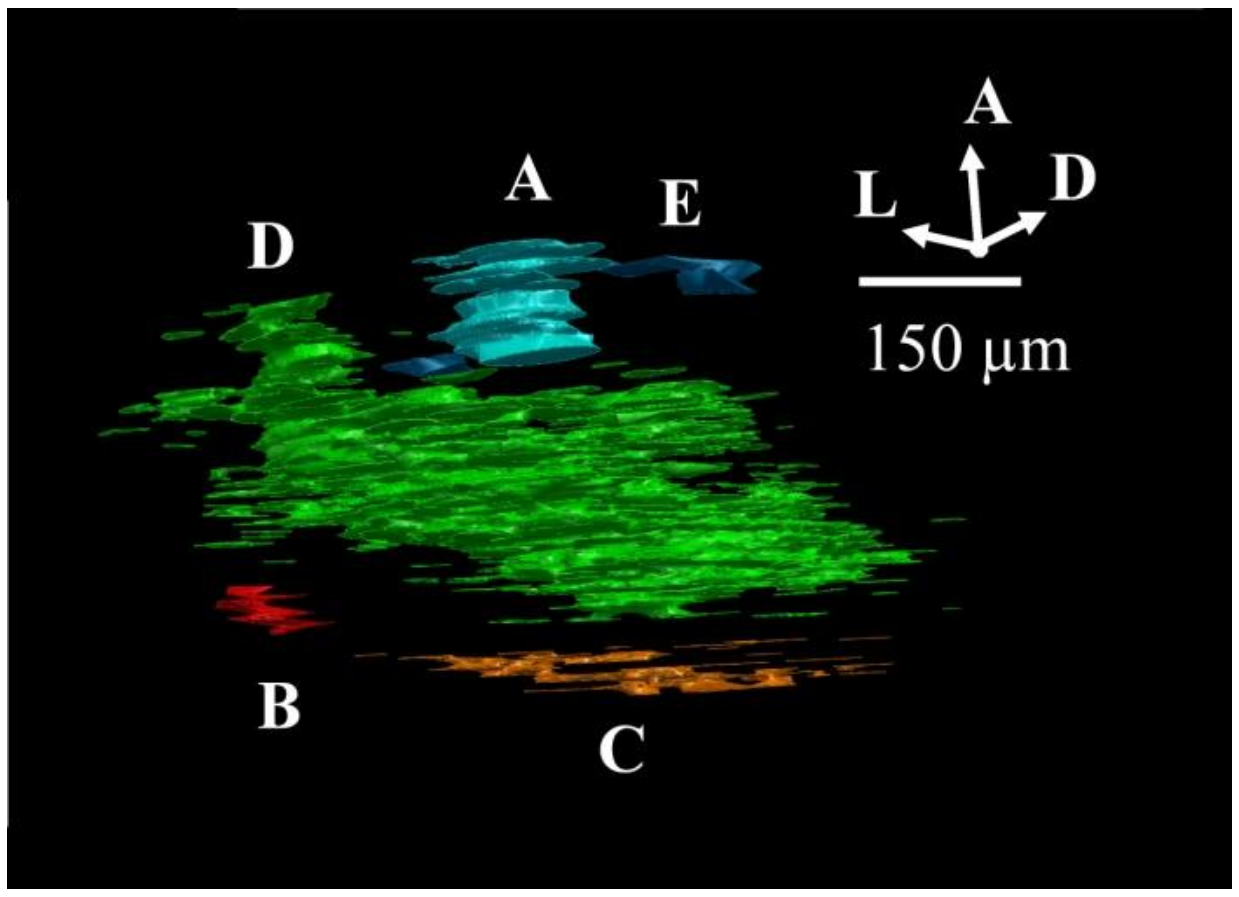

**Figure 2. Criteria for assigning 2D outlines to 3D subdomains. A.** Example of a “dense subdomain”, defined as an anterior-posterior z-stack of ipsilateral outlined regions, each of which are densely populated by ipsilateral termini (Figure 1 **C**), and whose outlines in neighboring sections overlap at least 10%. **B.** Example of a “sparsely-populated” subdomain, defined as a stack of outlines of sparse ipsilateral termini (Figure 2 **D**) in which outlines in neighboring sections overlap by at least 1%. **C.** Example of a “diffuse subdomain,” in which the outlines overlap less than 1% in the anterior-posterior (z) axis, yet the entire group of outlines are close to one another relative to the distance to the closest neighboring subdomain. **E.** Example of “stray outlines,” defined as outlines which neither overlapped with outlines in other sections, nor formed part of a group meeting criteria of a diffuse subdomain; these outlines were assigned to the closest subdomain in the dLGN, in this case, **A**.

Finally, in a few cases, stray outlines displayed some properties of a distinct subdomain but did not meet all criteria. In this case, as illustrated in Figure 2, **E**, these outlines were therefore grouped with closest subdomain (Figure 2, subdomain **A**).

For three of the specimens (Rats 5-7), we also imaged and traced the retinorecipient ventral lateral geniculate (vLGN, including the Inter Geniculate Leaflet or IGL), and the retinorecipient layer of the optic tectum for volume comparisons.

## Calculation of volumes

To estimate the volumes of structures on the basis of traced outlines, we used the method of Sackett et al. (1989) as implemented by Najdzion et al. (2009). For the *n*th section, the sub-volume $V_n$ is given by:

$$V_n = \frac{distance\ between\ sections}{3} \times (a_n + a_{n+1} + \sqrt{a_n} \times a_{n+1}) \quad \textit{Equation 1}$$

where $a_n$ is the area of the cross section through the structure of interest based on the traced outlines. The sub-volumes of the extreme sections (end poles) were estimated as:

$$V_n = \frac{distance\ between\ sections}{3} \times a_n \quad \textit{Equation 2}$$

The total volume of the structure $V_0$ is estimated by the sum of the sub-volumes throughout the region of interest:

$$V_0 = \sum V_n \qquad \textit{Equation 3}$$

This method was used to estimate the volume of the dLGN, the vLGN/IGL, and the SC (see Table 1). Due to the fragmented structure of the ipsilateral projection zones, we were not able to estimate the volume of the ipsilateral and contralateral projection zones by this method. Instead we compared the areas of ipsilateral and contralateral projection zones over all traced sections.

**Image Preprocessing for Analysis of Segregation**

To determine the extent to which termini from the two eyes segregate or overlap, the relative fluorescence was quantified at each location in the image. For this analysis images were preprocessed as follows. First, all fluorescent images throughout the brains were masked using the Neurolucida outlines in order to contain the dLGN only. Second, each image was corrected for bleedthrough fluorescence (crossover), which exists due to the overlap of the spectral profiles of the AlexaFlour 488 and 594 dyes. Specifically, very strong emission from the green fluorophor can be weakly captured by the red filter cube. To remove the resulting artifact, we identified all pixels containing 95% of the maximum green staining, and set the red intensities at those locations to zero. Images were then visually inspected for successful artifact removal. Third, we thresholded the images to remove background fluorescence. The threshold for each section was set to 99 to 99.9 percentile of the intensity values taken from other regions of the same brain section known to contain RGC fibers of passage but not termini. Thresholded images were visually inspected to confirm that fibers of passage in the dLGN were removed while termini were spared (Figure 3). Based on this inspection, the percentile cutoff was manually adjusted for each brain as needed, but then the percentile was held constant for all sections from that brain. After background subtraction, each of the two fluorescent channels was normalized to the maximum intensity of that channel in all sections of that brain.

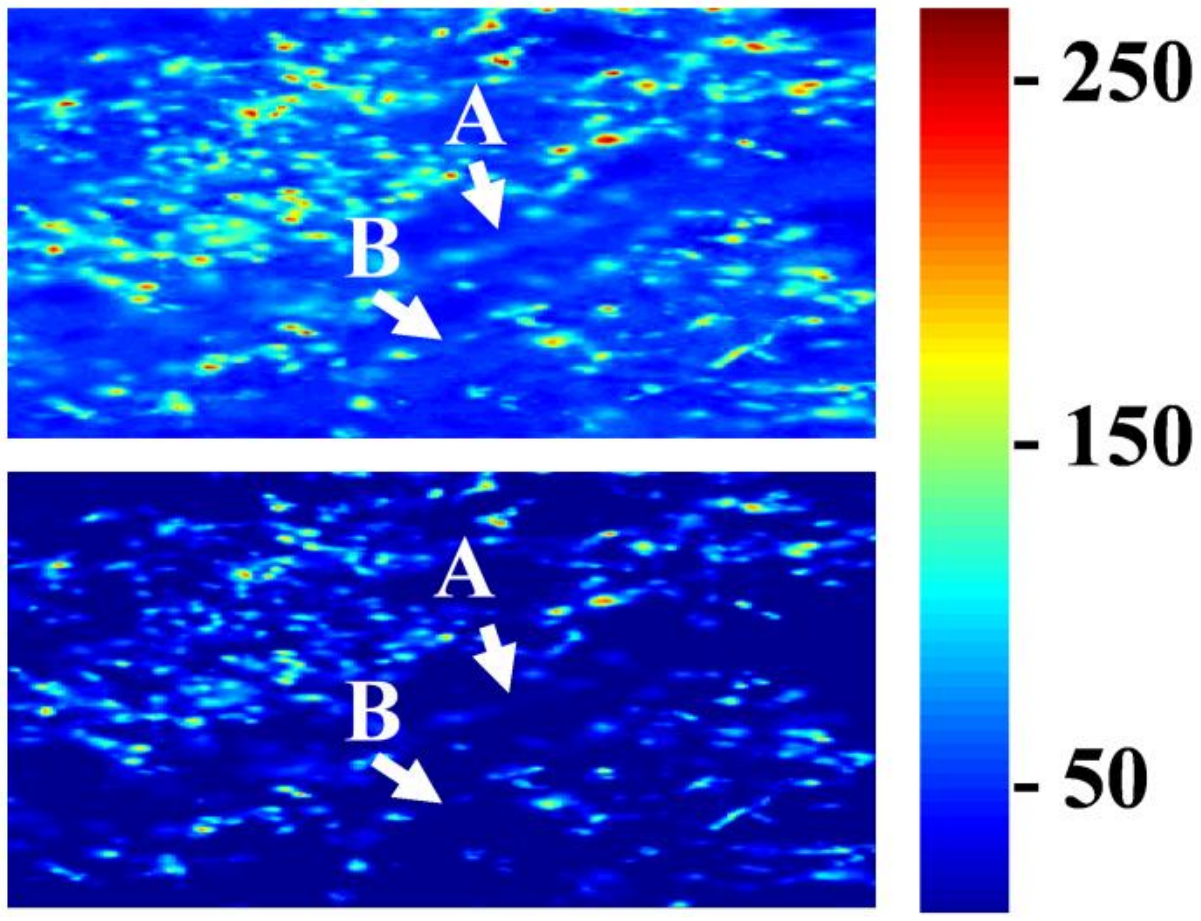

**Figure 3. Subtraction of background fluorescence from dLGN images.** Close up of an example image before background subtraction (top panel) and after background has been removed (lower panel). Pseudo-color indicates staining intensity, normalized to the range 0-255. Background threshold was chosen such that fibers of passage (**A**, **B**) were removed. This pre-processing step was used only for the segregation analysis (Figures 10-13), and ensured that only retinal termini contribute to calculations of binocularity.

**Measure of segregation of inputs by eye of origin**
To determine the overlap or segregation of ipsilateral and contralateral terminals, we used the analysis method of Torburg and Feller (2004), implemented in custom MATLAB code. For each pixel in the masked image we computed an *R* value:

$$R = \log\left(\frac{intensity\ of\ ipsilateral\ staining\ of\ pixel}{intensity\ of\ contralateral\ staining\ of\ pixel}\right)$$

*Equation 4*

Although *R* is continuous in value, for the purpose of summarizing results we classified an LGN location as 'monocular' when staining from the non-dominant eye was <1% that of the dominant eye, corresponding to an *R* value >2 or <-2. Thus any location with 1% or more contribution from the non-dominant eye (-2 <= *R* <= 2) was classified as binocular. This criterion is meant to be stringent with regard to our claim of strict segregation.

We observed uniform staining across the entire dLGN and optic tectum for Rats 1, 2, and 3, indicating complete filling of RGCs across the retina. In Rat 4, however, uneven and weak staining was observed in both retinal targets, indicating uneven filling of RGCs. It was still possible to visualize terminals clearly enough to manually outline the dLGN and termination zones by eye of origin, but this specimen did not pass the criterion for input segregation analysis, which depends on comparison of staining intensity.

**Control for low or unequal staining intensities**
One possible confound to our analysis of segregation is that in many cases, staining intensities of one or more of the fluorescent CTBs was weak. While CTB is known for complete filling of RGCs (Matteu et al., 2003; Rainer et al., 1996; Angelucci et al., 1996), it is also known for its frequently low-intensity fluorescence as well as degradation over time (Angelucci et al., 1996). For the analysis of segregation we only included the three binocularly-injected subjects in which staining was uniform throughout the major retino-recipient zones (see above). Nevertheless staining of the two fluorophores was generally unequal, with the red channel staining more weakly. Each channel was normalized to its own peak staining prior to the analysis of segregation (see above). Nevertheless, when staining is weak, it is possible that after subtracting background and fiber-of-passage fluorescence, some signal from the RGCs might have been missed. This could have biased the binocularity conclusions in favor of segregated eye inputs.

Therefore as an additional control, we analyzed the *R*-value distributions for each dLGN separately. In one dLGN the weaker stain represents the contralateral projection while in the other dLGN weaker stain represents the ipsilateral projection. *R*-distributions with red-stained contralateral input (Figure 4 **A**) showed far fewer pixels classified as contralateral monocular, compared with the dLGN with green-stained contralateral input (Figure 4 **B**). Despite this asymmetry, both samples support our main conclusion that few locations in the dLGN have equal staining (*R* ~=0), and a minority of locations have binocular staining (-2 < *R* < 2).

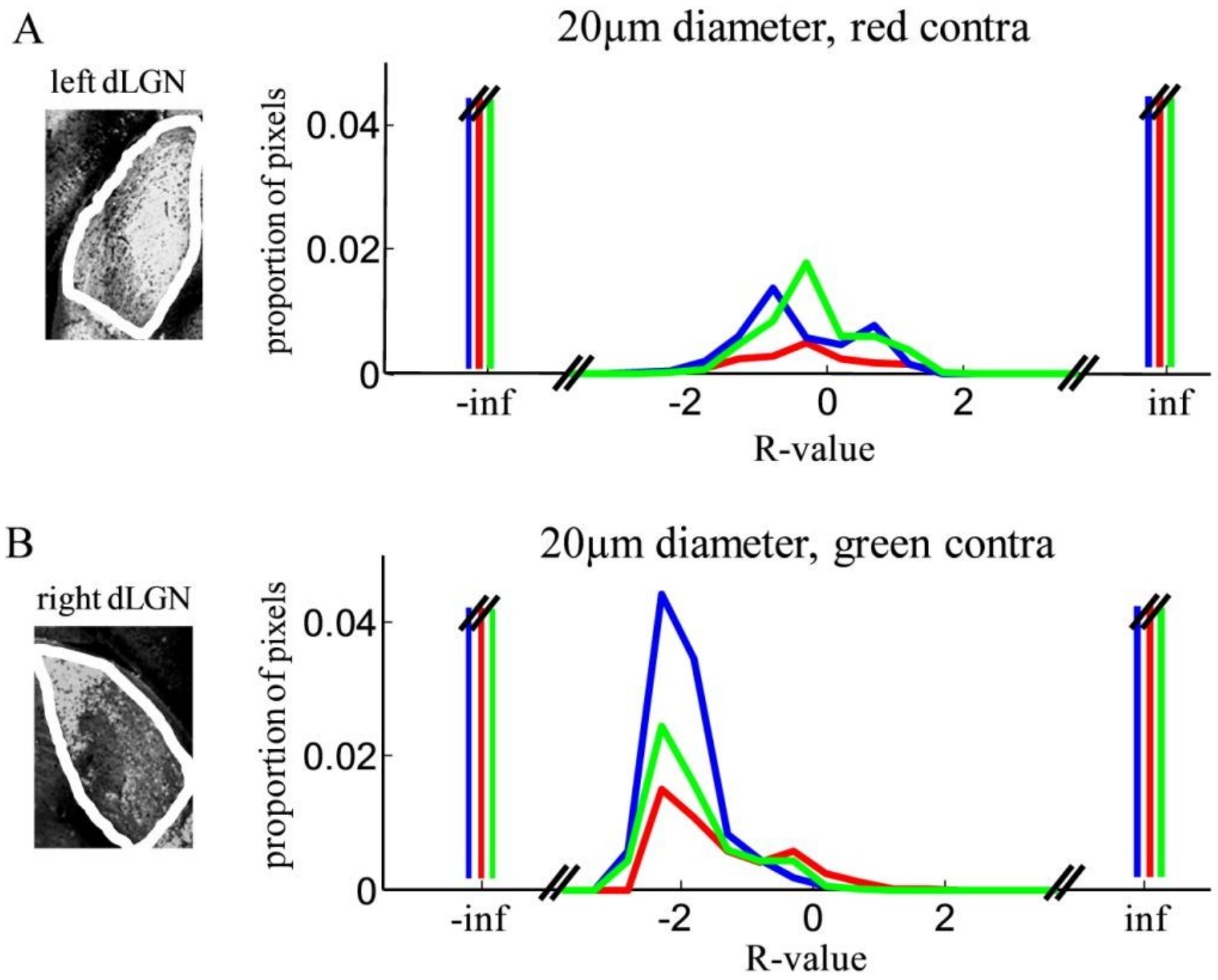

**Figure 4. Robustness of segregation analysis to unequal staining in the two eyes.** Eye-of-origin segregation was assessed by the log of the ratio of the intensity of ipsilateral to contralateral staining (*R*-value, see Methods). Here we show the same data as Figure 10, separated out based on hemisphere. **A.** *R*-value distributions of left dLGN samples (3 subjects), in which contralateral termini were stained red. **B.** *R*-value distributions of right dLGN samples, in which contralateral termini were stained green.

# RESULTS

## Imaging Retinal Termini

We injected fluorescently-conjugated CTB binocularly in four male Long-Evans rats in order to label retinal termini. Brains were later perfused and the region containing the dLGN sliced coronally into 25µm thick sections and imaged using a Nanozoomer 2.0 HT (see Methods). The resulting images are wide field (multiple entire sections contained in a single scanned image) and high resolution (0.46$\mu m^2$/pixel). All inputs from the left eye fluoresced green (488 nm), and all inputs from the right eye fluoresced red (594 nm).

A representative coronal section through the dLGN is shown in Figure 5. Viewed at moderate magnification, both left and right dLGN are visible, along with other retinorecipient structures in the subcortex (Figure 5 **A**). The right dLGN is shown at higher magnification in Figure 5 **B**. Stained retinal termini were also seen throughout other major retinorecipient targets, notably the optic tectum (Figure 5 **C**).

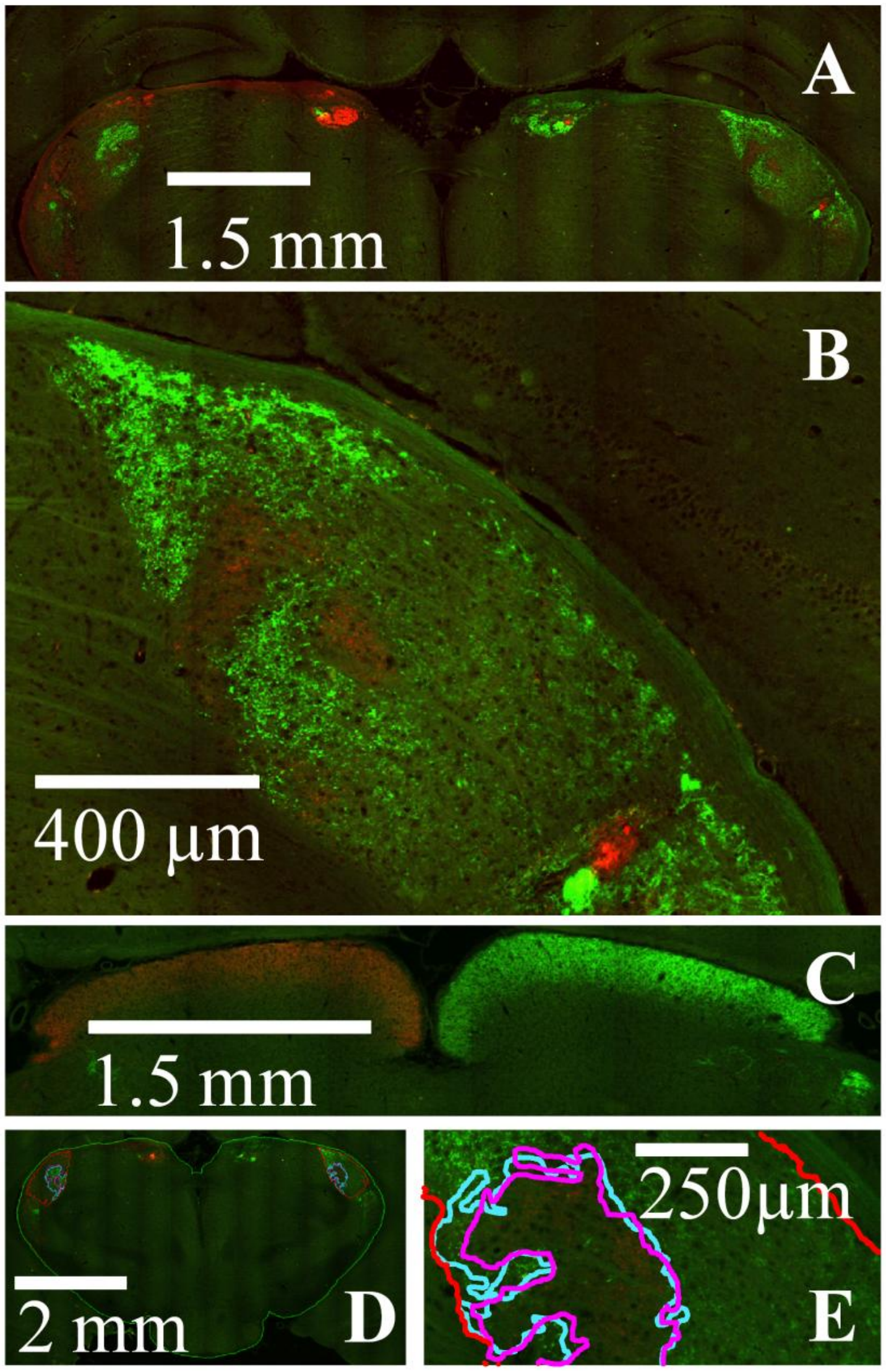

**Figure 5. Imaging of retinal termination zones.** High resolution wide-field images of retinal termini in subcortical targets. Alexa-Fleur 488nm-conjegated CTB (green) was injected in the left eye and AlexaFleur 594nm-conjugated CTB (red) in the right eye. **A.** A field of view within a coronal section from Rat 1, in which both dLGN as well as other retinal targets are visible. **B.** Expanded view from the section in **A**, showing the right dLGN. **C**. Field of view from a different coronal section from Rat 1, showing retinal projections to the optic tectum. **D**. Field of view from a different section showing manually traced outlines of the entire subcortex (*green*), entire dLGN (*red*), holes in the projections from the contralateral eye (*magenta*), and projections from the ipsilateral eye (*cyan*). **E**. Expanded view from **D** at the high resolution used to identify retinal termini during tracing, showing in the relationship between the contralateral hole and the ipsilateral projection outlines in this section.

Three additional rats were monocularly injected with non-conjugated CTB for analysis by light microscopy (not shown). Brains of these subjects were sliced coronally at 30µm, termini stained with DAB, cell bodies counter-stained with Geimsa, and digitally imaged by light microscopy using the Aperio Scanscope (see Methods). The resulting images are wide field (multiple entire sections contained in a single scanned image) and high resolution (0.5$\mu m^2$/pixel). In these specimens we sectioned and imaged the entire brain, enabling volumetric analysis of additional retinal targets.

### Tracing retinal projections

Contralateral and ipsilateral retinal projection zones were traced in the dLGN on both sides of the brain (Figure 5 **D**). Tracing was performed using the highest available magnification (Figure 5 **E**) according to defined procedures and criteria (see Methods; Figure 1).

We traced every section through the lateral geniculate nucleus (LGN) and surrounding brain for one binocularly injected, fluorescently labeled sample. We traced every fourth section through the lateral geniculate nucleus (LGN) and surrounding brain for an additional 3 binocularly injected, fluorescently labeled samples and 3 monocularly injected, non-fluorescent samples.

The resulting outlines were used to determine the volume of the dLGN and other structures (Table 1), to determine the three-dimensional structure of ipsilateral subdomains in the dLGN (Figures 6-9), and to determine the extent of overlap of the ipsilateral and contralateral projections to the dLGN (Figures 10-13).

### Volume of the dLGN

In our samples the average volume of one dLGN was 1.58 $mm^3$±0.094 $mm^3$ (mean±SD, n=11 dLGN nuclei from 7 rats; Table 1). The dLGN comprised 70.0% (±3.0%, n=3) of the total RGC-recipient geniculate volume, which includes the vLGN, the intergeniculate leaflet (IGL), and the dLGN. The volume of the dLGN was 40.4% (±1.0%, n=3) that of the optic tectum.

| Subject # | left dLGN | right dLGN | left IGL/vLGN | left optic tectum |
|---|---|---|---|---|
| 1 | 1.76 $mm^3$ | 1.56 $mm^3$ | -- | -- |
| 2 | 1.63 $mm^3$ | 1.59 $mm^3$ | -- | -- |
| 3 | 1.55 $mm^3$ | 1.59 $mm^3$ | -- | -- |
| 4 | 1.66 $mm^3$ | 1.63 $mm^3$ | -- | -- |
| 5 | 1.53 $mm^3$ | -- | 0.55 $mm^3$ | 3.70 $mm^3$ |
| 6 | 1.46 $mm^3$ | -- | 0.69 $mm^3$ | 3.71 $mm^3$ |
| 7 | 1.42 $mm^3$ | -- | 0.65 $mm^3$ | 3.51 $mm^3$ |

**TABLE 1: Volumes of retinorecpient structures.** Volumes of retinorecipient structures were determined from traced outlines of retinal termination zones (see Methods). Subjects 1, 2, 3, and 4 were binocularly injected, and therefore the volume of the dorsal lateral geniculate nucleus (dLGN) could be measured on both sides of the brain. Subjects 5, 6 and 7 were monocularly injected, so volumes could only be measured on the side contralateral to the injection. In these subjects, however, the entire brain was sectioned and imaged so that the optic tectum and the ventral lateral geniculate nucleus (vLGN) / intergeniculate leaflet (IGL) could also be reconstructed.

**Putative ipsilateral subdomains within the dLGN**

The area of the dLGN receiving ipsilateral input was 12.08±1.82% (mean±SD, n=8) of the retinorecipient dLGN. In many sections, we observed two or more spatially separated zones of ipsilateral termini (e.g., Figure 5 **B**), rather than the single compact termination zone expected based on the literature. These ipsilateral zones were well-aligned holes in the contralateral projections (Figure 5 **E**). To determine whether these were part of a single connected three-dimensional (3D) ipsilateral recipient zone, we reconstructed the dLGN and its retinal termination zones in 3D for all four binocularly-injected subjects.

The contralateral and ipsilateral projection volumes of one subject are shown in Figure 6 **A**. We found several spatially separated subdomains of ipsilateral termini within each dLGN (Figure 6 **B**), based on criteria that favored lumping over splitting (see Methods).

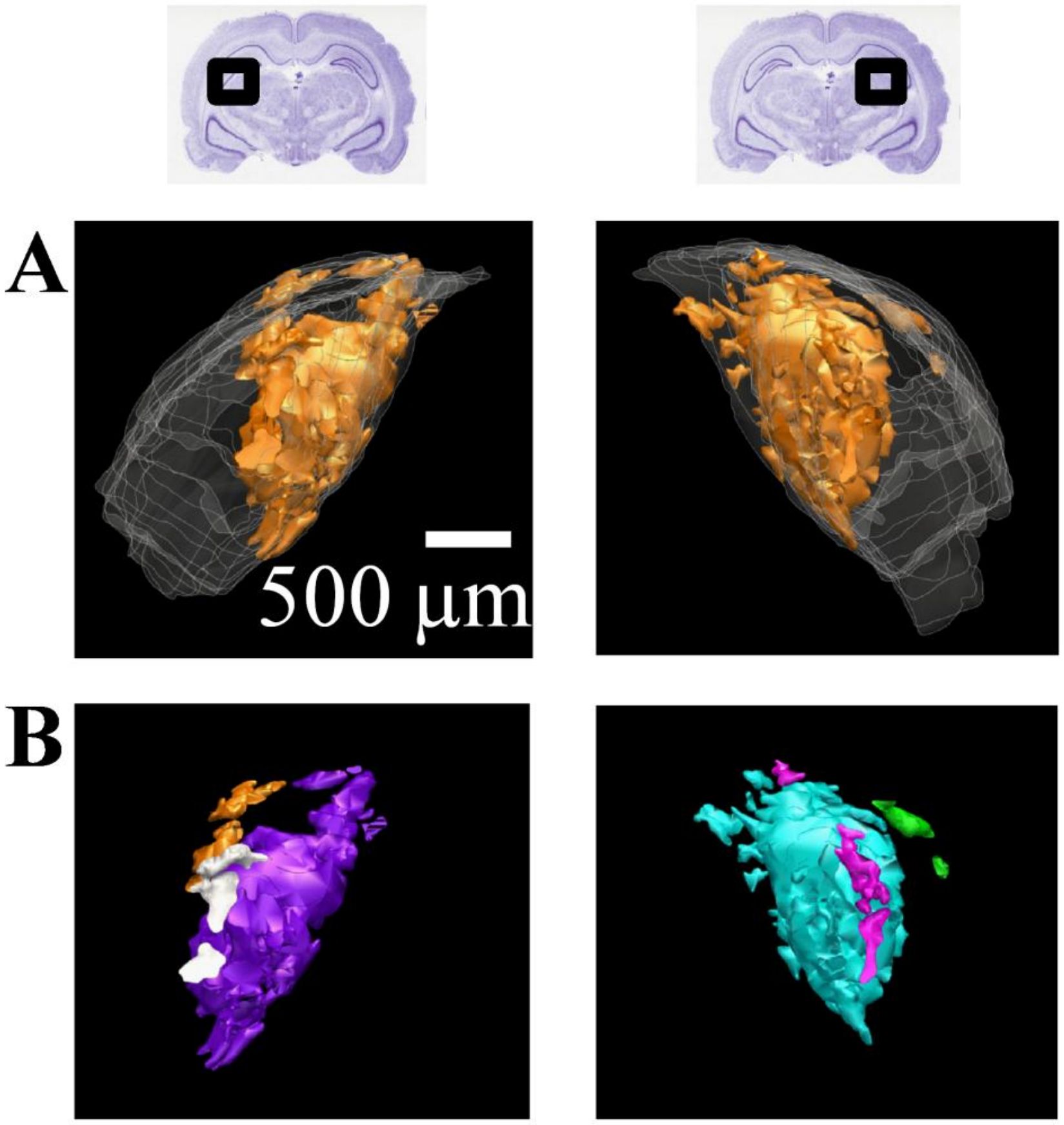

**Figure 6. 3D reconstructions of ipsilateral subdomains within the dLGN of one rat.** A 3D reconstruction of the left (left) and right dLGN (right) of Rat 1, as schematically indicated by inset Nissl-stained sections above. For this subject only, every section through the brain was traced for higher resolution in the z-axis. **A.** All the ipsilateral-recipient subdomain volumes shown in *orange*; the outline of the entire dLGN shown in translucent *white*. **B.** The same ipsilateral subdomains shown in **A**, but each spatially separate subdomain is indicated in a different color. No specific correspondence between particular left and right subdomains is claimed, so different colors are used for the two hemispheres.

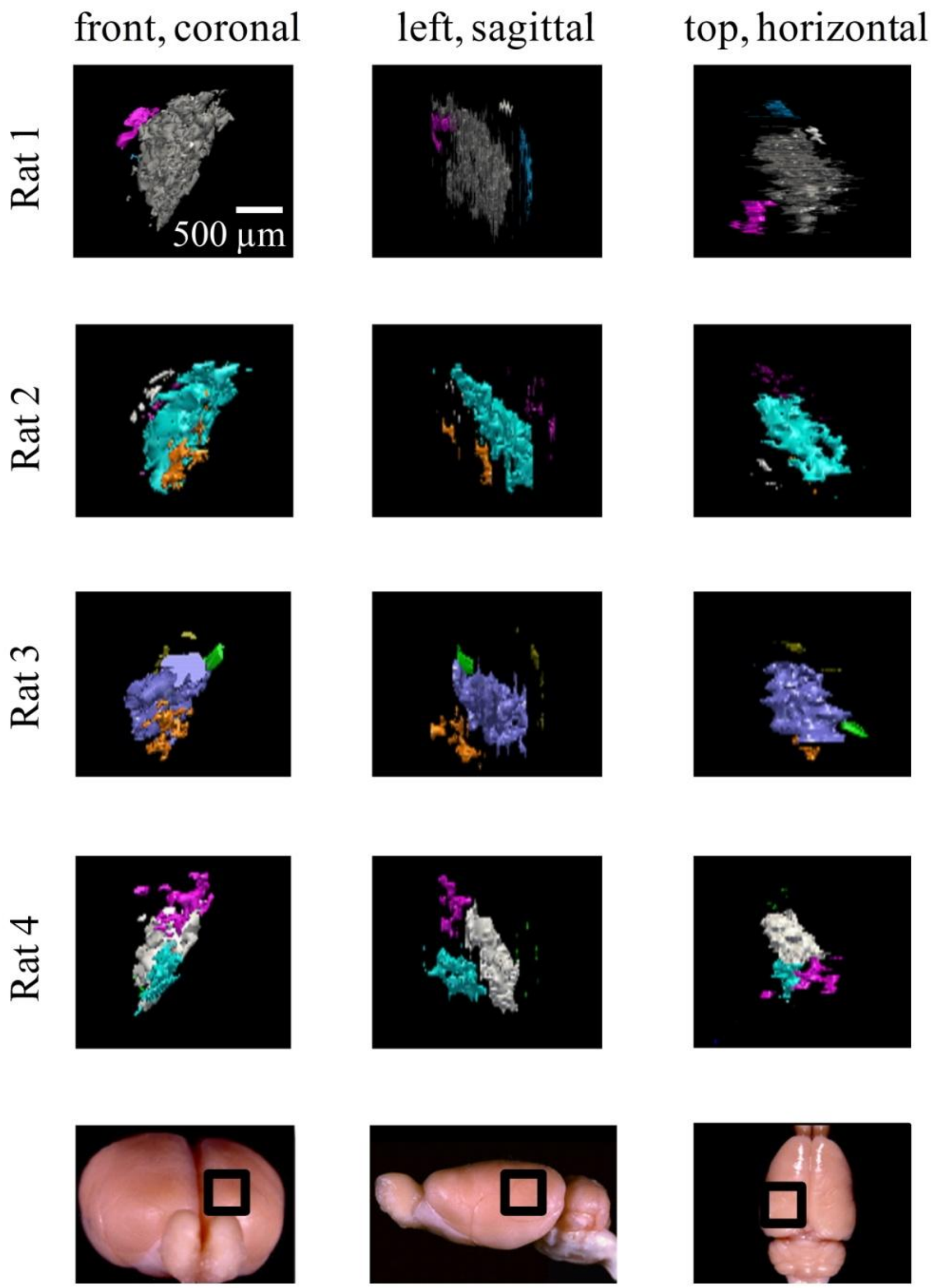

**Figure 7. Ipsilateral projections to the left dLGN of four subjects.** All the ipsilateral-recipient subdomain volumes in the left dLGN of each binoncularly-injected subject, with each spatially separate ipsilateral subdomain indicated in a different color. Each 3D reconstruction is shown from three different vantage points: the top, front, and side view. The vantage point is illustrated at the bottom, with the dLGN position marked by a black square on a whole-brain icon. No specific correspondence between particular subdomains of different subjects, nor across hemispsheres, within subjects is claimed, so different colors are used for each subject and hemisphere.

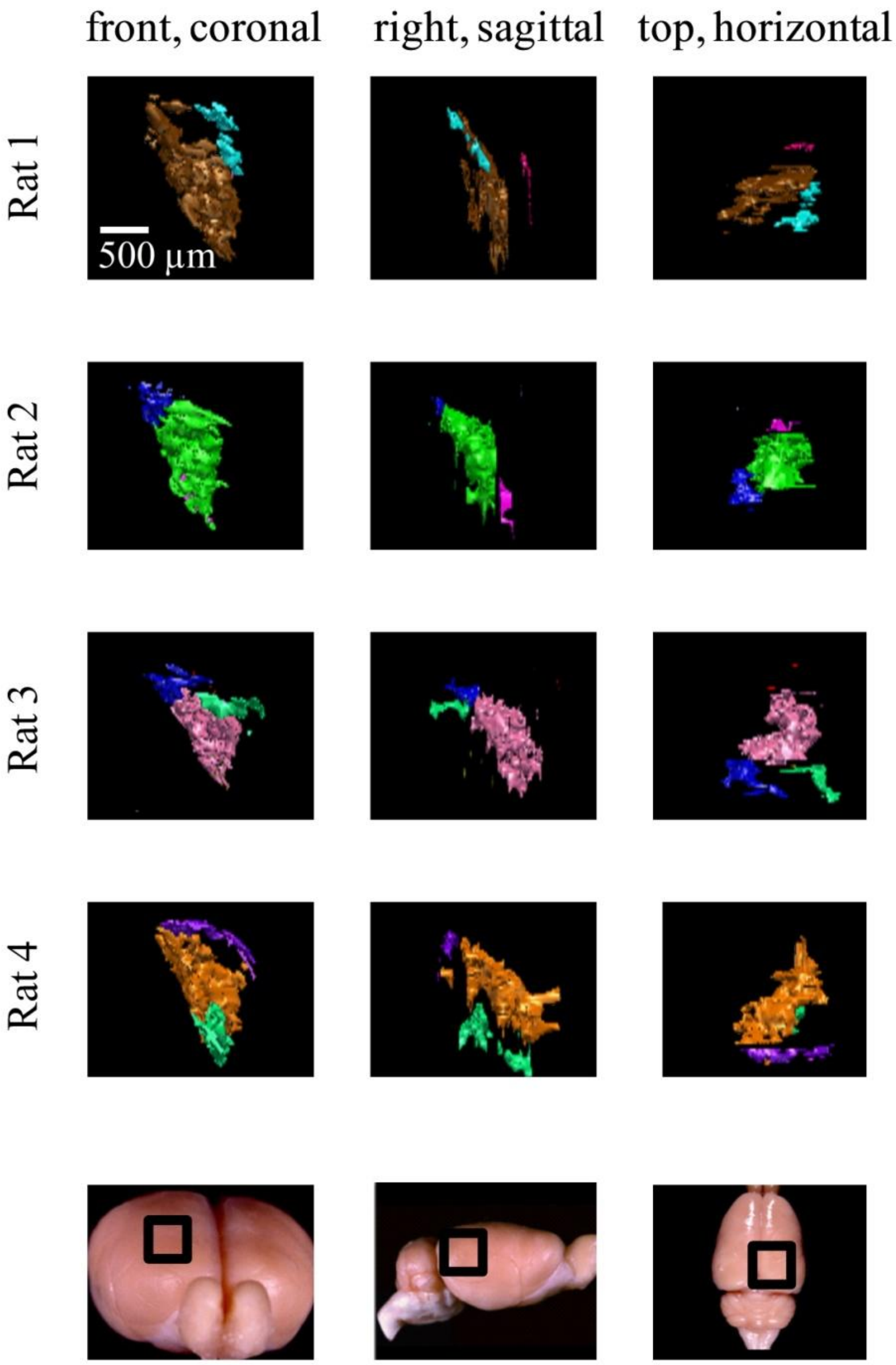

**Figure 8. Ipsilateral projections to the right dLGN of four subjects.** All the ipsilateral-recipient subdomain volumes in the right dLGN of each binoncularly-injected subject, with each spatially separate ipsilateral subdomain indicated in a different color. Each 3D reconstruction is shown from three different vantage points: the top, front, and side view. The vantage point is illustrated at the bottom, with the dLGN position marked by a black square on a whole-brain icon. No specific correspondence between particular subdomains of different subjects, nor across hemispsheres, within subjects is claimed, so different colors are used for each subject and hemisphere.

Similar results were found in all four subjects (left hemisphere dLGN, Figure 7; right hemisphere dLGN, Figure 8). In general, three categories of ipsilateral subdomains were found: a dorsal-medial, a ventral-rostral, and a larger central region. We show the reconstructions for both hemispheres of all subjects from three perspectives, to allow direct inspection of the degree of bilateral symmetry as well as inter-subject variation.

While the number of these subdomains (Figure 9) and their exact locations varied from animal to animal and even between hemispheres in the same animal (Figures 7 and 8), the approximate locations of these ipsilateral subdomains remained generally consistent. From these data we conclude that the dLGN of the pigmented rat typically contains multiple spatially separated ipsilateral projection zones, and not one single zone as described previously.

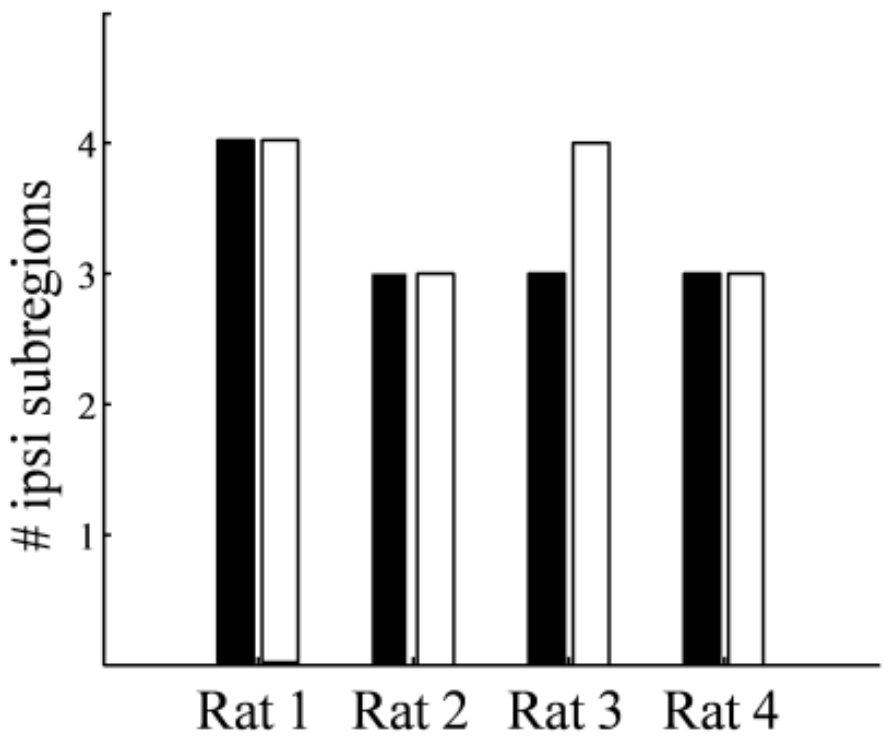

**Figure 9. Number of spatially separated ipsilateral-recipient subdomains in the dLGN.** For each binocularly injected subject (Rat 1 – Rat 4), the number of identified subdomains in the left dLGN (*black*) and right dLGN (*white*) as determined by 3D reconstruction.

**Spatial segregation of retinal termination zones within the dLGN.**
We found little overlap in the traced outlines of ipsilateral and contralateral projection zones in the dLGN, consistent with strict segregation by eye of origin as described in other mammals. Considering that binocular responses have been reported in the literature (Greive, 2005), however, the degree of segregation has been questioned. To address this further, we used a method introduced by Torborg et al. (2008) to measure segregation using the relative intensity of staining of retinal termini originating from the two eyes. We computed for each location in the dLGN an index of binocularity $R$, defined as the log of the ratio of ipsilateral to contralateral staining (see Methods, Equation 4). The index $R$ has a negative value when contralateral inputs are stronger, a positive value when ipsilateral inputs are stronger, and is 0 when the normalized intensity originating from the two are equal.

The result of this analysis will depend critically on the spatial sampling diameter over which intensity is measured. In the limit of analyzing single submicron pixels, each "location" is smaller than a single retinal terminal, so contributions at that spatial scale will be monocular, even in a binocular structure with completely mixed, unsegregated inputs. In the limit of large sampling diameter, a single "location" could include the entire dLGN, and contributions will be binocular even for a structure with well-segregated inputs. In general, we expect binocularity to increase with sampling diameter. The choice of sampling diameter is somewhat arbitrary, so we present results as a function of this variable (Figure 10-13). Our findings are robust to choice of this parameter.

At a sampling diameter of 1μm, over 90% of locations in the dLGN have monocular input (no measurable staining originating from the other eye). Some locations, however, have measurable staining in both channels, implying at least some retinal termini from each eye of origin (Figure 10 **A**). We operationally defined locations with $-2 < R < 2$ as "binocular"; locations with $R <= -2$ as monocular and contralateral; and locations with $R >= 2$ or greater as monocular and ipsilateral. This classification is meant to be stringent relative to a claim of segregation: if even 1% of the staining originates from the non-dominant eye the location is considered binocular, even though no relay cell may in fact sample from termini of both eyes at that location.

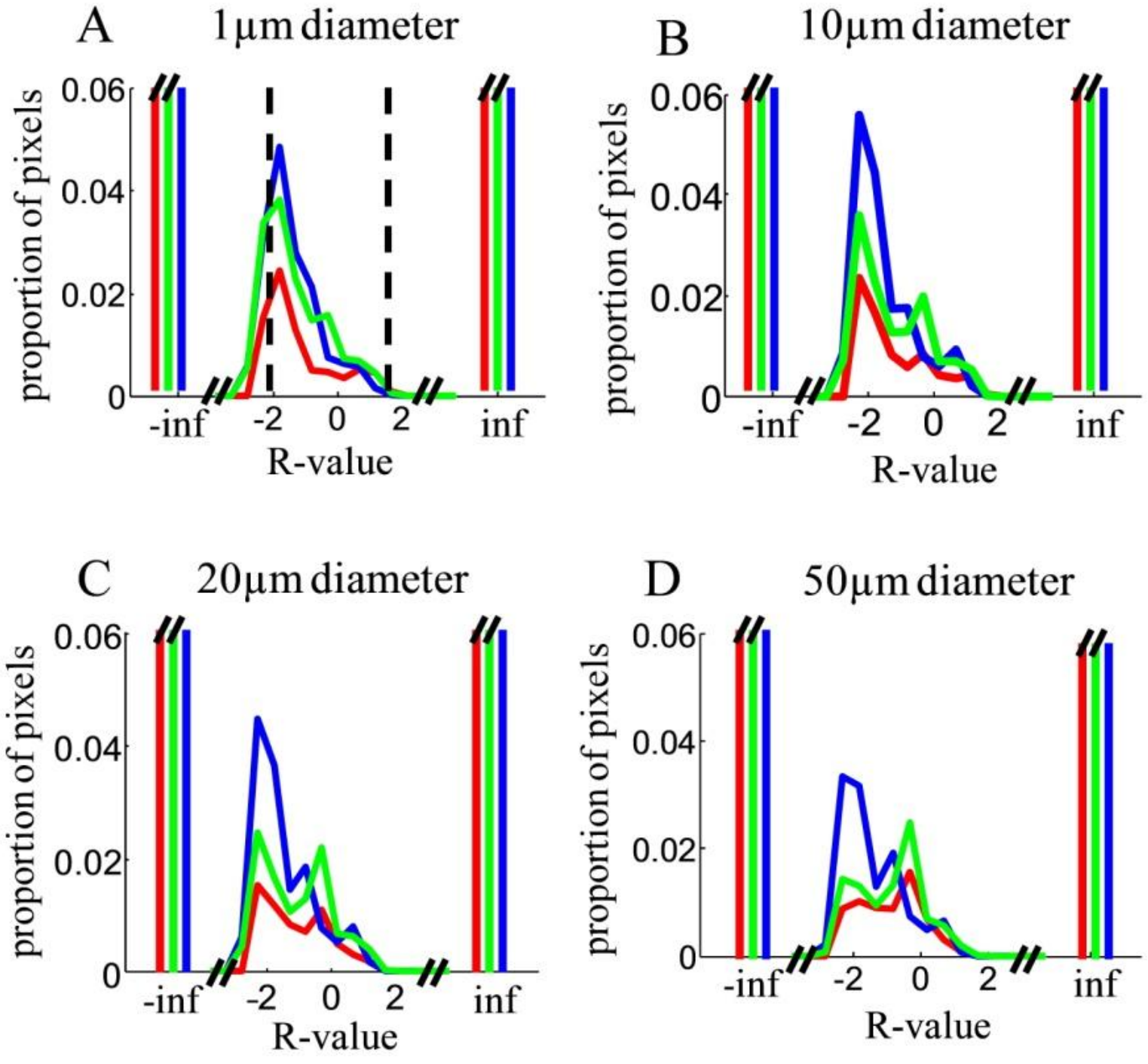

**Figure 10. Relative contribution of inputs from the two eyes. A**. Distribution over locations in the image of the *R*-value, i.e. the log of the ratio of ipsilateral to contralateral staining intensity (Equation 4), analyzed in a sampling diameter of 1μm. Distributions were computed over the entire area of the dLGN in both hemispheres in all analyzed sections. Results are shown for Rat 1 (*green*), Rat 2 (*red*) and Rat 3 (*blue*). The cutoff values for our operational definition of "binocular" ($-2 < R < 2$) are indicated by vertical *dashed lines*. Distributions were normalized such that the histograms sum to 1. The y-axis is greatly expanded to show the shape of the distribution for the small minority of pixels with detectable staining from both eyes; the proportion of monocular pixels (R = or R = -) is far off-scale. **B-D**. As in **A,** but analyzed in a sampling diameter of 5μm, 10μm, and 20μm respectively.

Across a wide range of sampling diameters (1μm -50μm, Figure 10 **A-D**), most of the dLGN locations classified as binocular have stronger input from the contralateral eye (peak near $R$=-2, corresponding to 100:1 excess of contralateral staining). The contralateral contribution was stronger (peaked at $R$<0) regardless of whether the contralateral eye was the weaker or the stronger staining (see Methods, Figure 4). Additional smaller peaks were often observed near $R$=0 (equal contribution) and near $R$=0.5-1 (ipsilateral dominating by three- to ten-fold).

In principle, an $R$ value near 0 (staining ratio near 1) could arise from extremely weakly stained locations in both channels; the ratios of very small numbers would not be reliable due to noise. The joint histogram of staining intensities in the two eyes (Figure 11) reveals, however, that most locations classified as binocular arose from locations with clearly measurable staining in both eyes (log intensities >2 in both channels, corresponding to >=1% of maximum intensity in each channel). Most dLGN locations classified as monocular by our criteria had no detectable staining (intensities $<10^{-6}$) in the non-dominant eye (Figure 11 **A**, compare left vs right panels). As the sampling diameter increased from 1µm to 20µm (Figure 11 **B-E**), so does the number of locations in the dLGN that show equal contributions from the two eyes (density along x=y diagonal). Yet up to a sampling diameter of 20µm, most binocular pixels were dominated by either contralateral or ipsilateral input (off-diagonal density).

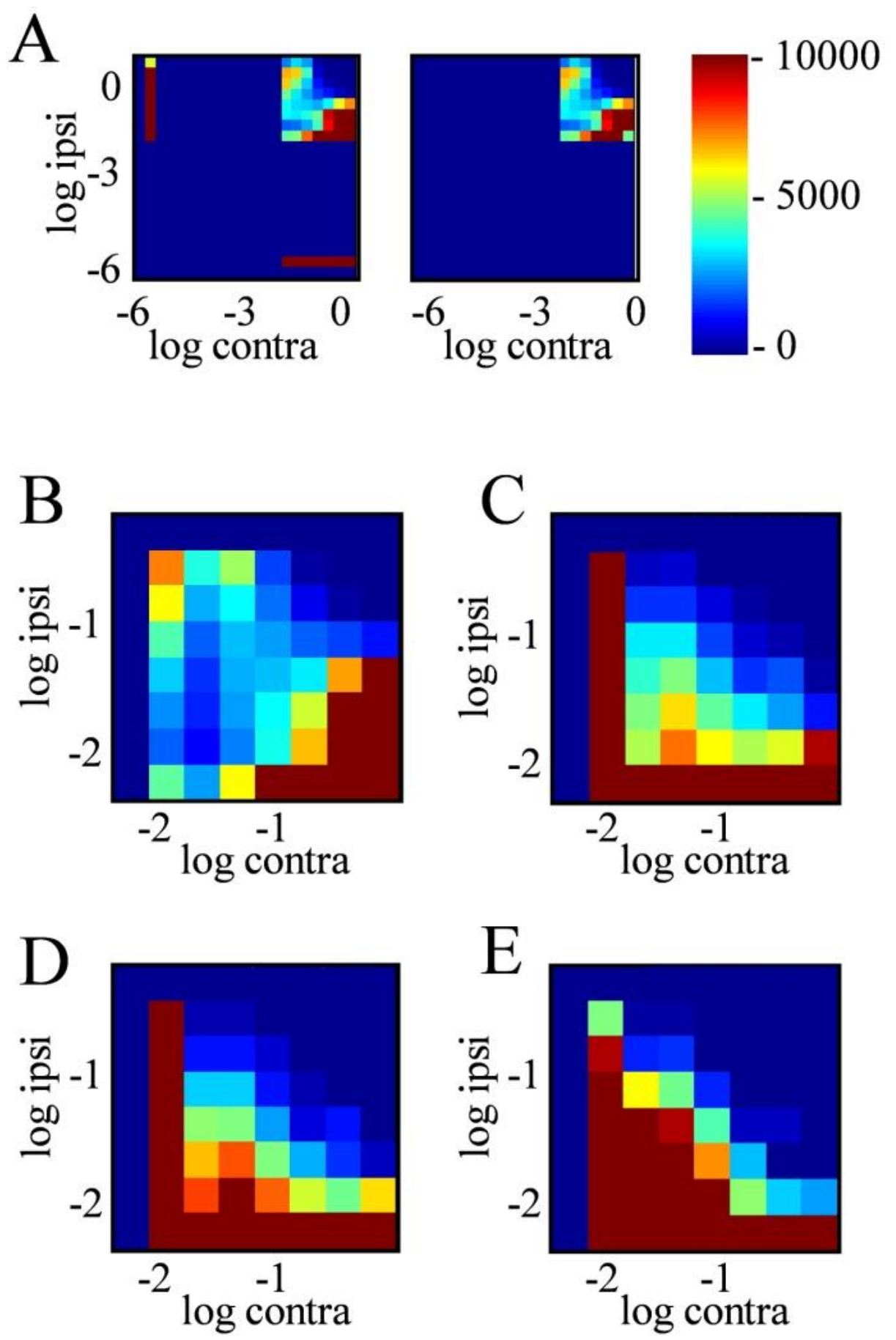

**Figure 11. 2D histograms of ipsilateral and contralateral input strength. A.** Joint probability distribution of log contra staining intensity (x-axis) vs. log ipsi staining intensity (y-axis) over positions in the dLGN, where color indicates the number of locations in the dLGN with these staining intensities. Data are from the entire dLGN, both hemispheres, all sections of Rat 1, using a sampling diameter 20µm. Intensity was discretized in bins of 0.3 log units². The *left subpanel* includes all positions in dLGN; the *right subpanel* includes only "binocular" locations (-2 < *R* < 2). **B-E.** Expanded view of the joint probability distribution for binocular locations only, analyzed in a spatial sampling diameter of 1, 5, 10, and 20µm respectively.

The spatial distribution of *R*-value reveals that most dLGN positions classified as binocular lie at the boundaries between monocular regions (Figure 12 **A-C**). At a sampling diameter of 20µm, for example, only 5% of pixels in the section shown were classified as binocular, and most of these fell along the boundaries between ipsilateral- and contralateral- recipient regions (Figure 12 **C**). The percentage of pixels classified as binocular is shown as a function of sampling diameter for all three subjects (Figure 13). At a sampling diameter of 20 µm, between 90-96% of all positions were classified as strictly monocular.

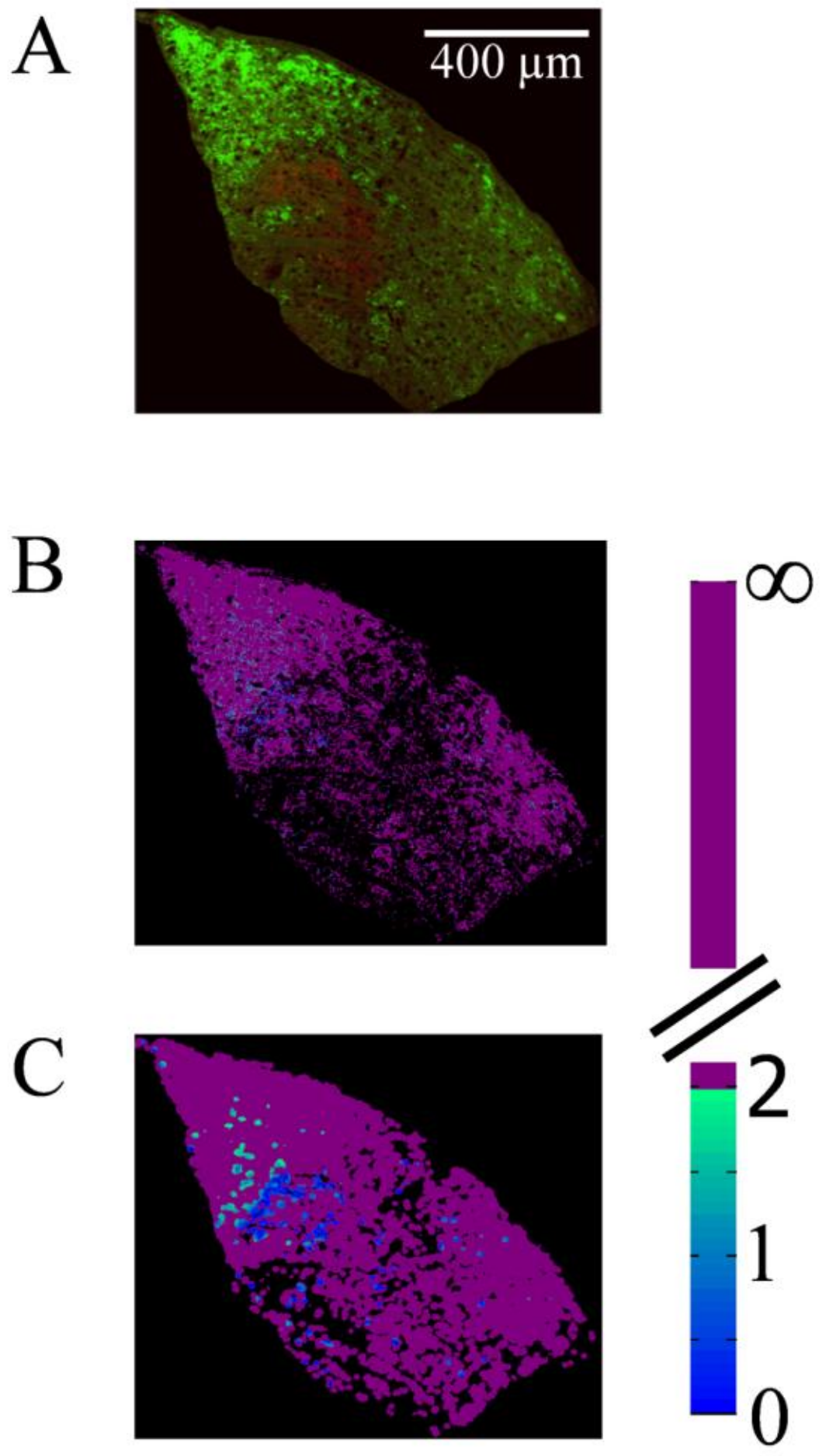

**Figure 12. Spatial distribution of *R*-values. A**. Merged image of the contralateral (green) and ipsilateral (red) retinal termini, as labelled by the fluorescence of conjugated CTB tracers, in a representative slice through the dLGN. Brightness and contrast have been adjusted for illustration purposes; area outside the dLGN has been masked out. **B** and **C:** False-colored images of same section, where color indicates the absolute value of *R* (color scale at right), for a spatial sampling diameter of 1 µm (**B**) and 20um (**C**).

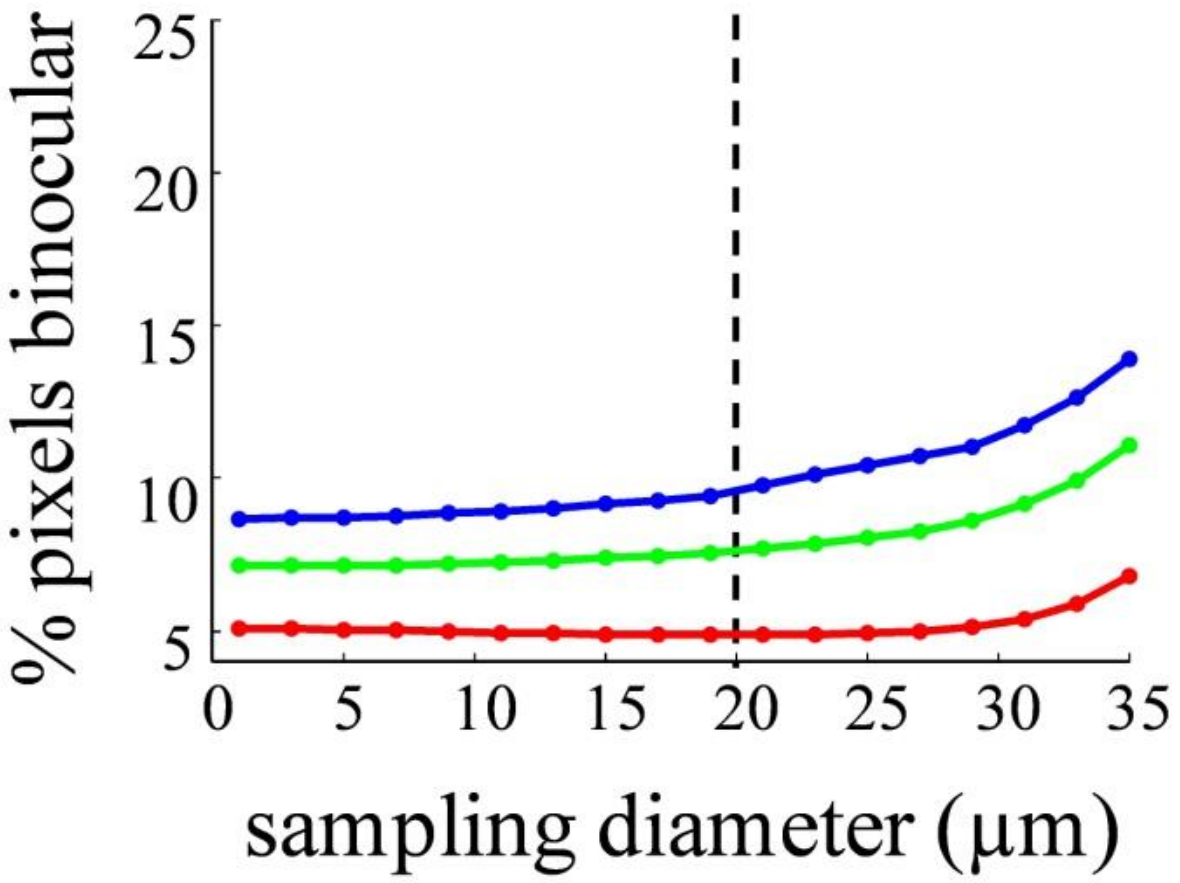

**Figure 13. Probability of an area containing terminals from both eyes depends on sampling diameter.** The percentage of binocular pixels (-2 < *R* < 2) as a function of the spatial sampling diameter, for Rat 1 (*green*), Rat 2 (*red*) and Rat 3 (*blue*). Our estimate of the diameter of a rat dLGN cell soma is shown in the *dashed line*.

# DISCUSSION

The data and analysis presented here confirm the basic findings of an earlier preliminary report (Discenza et al, 2008): retinal projections to the dLGN of the rat are well segregated by eye of origin, and the ipsilateral projections form multiple spatially separated subdomains.

**Volume of dLGN**

The volume of the dLGN relative to the entire retinorecipient thalamus (dLGN, IGL and vLGN) has been related to the visual sophistication of species. Najdzion et al. (2009) found that the contribution of the dLGN to the total LGN volume was 57% in the common shrew and 50% in the bank vole, which are both nocturnal and partially subterranean species. The relative size of the dLGN was considerably larger in the more visually dependent rabbit (64%) and fox (95%). Here we found the rat dLGN was 70 ± 3% of the LGN, placing it closer to the highly visual end of the spectrum of mammals (Table 1).

Braur et al. (1982) found that among 16 species, within a given order, those with a high level of neocorticalization also tended to have a high ratio of dLGN to vLGN volumes. In addition, ratios of dLGN to vLGN size were positively correlated with extent of dLGN lamination. The high ratio of dLGN to vLGN we found in the rat would be consistent with a laminated dLGN, despite the absence of obvious structural laminae. The entire retinorecipient thalamus was nevertheless smaller than the volume of the retinorecipient tectum (58 ± 1%, n=3).

**Multiple ipsilateral termination zones**

Ipsilateral projection zones comprised 12% of the area of the retinorecipient dLGN, consistent with the proportion of RGC crossover in the optic chiasm, as well as the percentage of binocular overlap in the rat's field of vision.

Rather than a single ipsilateral domain, three or four spatially separated subdomains of ipsilateral input were consistently found in each dLGN (Figure 9). These subdomains were typically seen dorsal-medially, centrally, and rostral-ventrally, though the exact locations and volumes were not well-conserved between subjects or even across hemispheres of the same brain. The presence of spatially separated ipsilateral subdomains raises the possibility of multiple interleaved ipsilateral and contralateral sublaminae.

In other mammals, sublaminae that represent parallel processing streams in the dLGN are sometimes but not always distinguishable in Nissl-stained sections on the basis of soma size and density. The putative subdomains we identified on the basis of termination zones appeared similar in these characteristics under Nissl stain; a quantitative analysis described elsewhere failed to find any statistical difference in soma size or density between different identified subdomains (Discenza, 2011). However, this doesn't preclude the existence of morphologically, functionally, or architecturally distinct subdomains, which may yet be revealed by other methods. In particular, we have not determined the retinal ganglion cell terminal morphology, which reveals hidden sublamination of the dLGN of other species (Major et. al., 2003).

Functionally distinct sublaminae are expected to receive input from distinct RGC subtypes. Several early studies found general differences in the anatomical types of RGCs that project to the `outer shell' vs. the `inner core' of the dLGN (Martin, 1986; see Reese, 1988). These studies found that type I (alpha) RGCs (cells with large somas and 3-6 primary branching dendrites) synapsed in the dLGN `inner core', type II (B) cells (small somas with short dendrites) synapsed throughout the nucleus, and type III (C) RGCs (cells with smaller somas and very long dendrites) were found in the `outer shell' only. But there are also at least a dozen functional subclasses of RGCs in the rat, each transmitting their own distinct information (Yonehara, 2009). Using new techniques to trace individual functional cell types, such as molecular tags (Marc and Jones, 2002) and genetic markers (Huberman et al., 2008, 2009), one could test whether the spatially separated ipsilateral domains receive projections from distinct RGC populations.

If the ipsilateral subdomains we describe here represent distinct functional laminae, they should contain separate retinotopic maps. Alternatively, if they represent a single spatially fragmented layer, they should jointly contain a single retinotopic map. The retinotopy of the rat dLGN has been described from physiological data (Reese and Jeffrey, 1983; Reese, 1988; Reese and Cowey, 1983). But given the small size and variable position of the ipsilateral subdomains, it would be necessary to fill and reconstruct the recorded cells to make any detailed comparison of retinotopy or physiological properties between subdomains. Retrograde labeling from V1 would provide valuable information about the retinotopic map(s) in the ipsilateral subdomains.

**Segregation of inputs by eye of origin**

Early studies reported that ipsilateral and contralateral retinal projections are segregated in the mature rat dLGN (Reese and Jeffrey., 1983; Reese and Cowey, 1983; Reese, 1988). One recent study, however, reported that up to 63% of relay cells in the rat dLGN respond to direct stimuli from either eye (Grieve, 2005), casting some doubt as to the degree of segregation of inputs from the two eyes.

While it is known that dLGN relay cell dendrites span nearly the entire nucleus (Gabbott et al., 1986) it has been shown that relay cells only synapse with RGC termini close to the soma (Hamos et al., 1985; for review, see Sherman and Guillery, 1996). We previously estimated the average cell diameter in the rat dLGN to be 20µm(Discenza 2011); others reported a maximum diameter of 15µm (Villena et al., 1997). If a dLGN relay cell samples retinal termini over a diameter of 20µm, only 5-10% of locations within the dLGN have access to termini from both eyes, and most of these locations fell along the borders between ipsilateral- and contralateral-recipient regions (Figures 12-13). LGN relay cells receive inputs from only 1-5 retinal ganglion cells (Levick et al., 1972); therefore even if they sampled uniformly within this radius, most would still be monocularly innervated. Therefore if binocular responses of dLGN relay cells are confirmed in the rat, they would more likely be explained by non-retinal inputs.

**Rat as a model system for vision**

Until relatively recently, rats have been regarded as largely non-visual animals, better known for their ability to use sense of smell and whisker-touch to navigate their environments (Hutson and Masterson, 1986; Hill and Best, 1981; Carvell and Simons, 1990; Maaswinkely and Whishaw, 1999; Kulvicius et al., 2008; Save et al., 2000). Yet despite their poor acuity and limited color vision (Jacobs, 2001; Prusky et al., 2002, for review see Burn, 2008), pigmented rats can learn and perform a wide range of visual tasks. In the laboratory setting, rats have demonstrated visuo-spatial learning and memory (Zoladek and Roberts, 1978; Morris, 1984), navigation (Holscher et al, 2004), and visual object detection and pattern discrimination (Thompson and Solomon, 1954; Zoccolan et al., 2009; Meier Flister and Reinagel 2011; Clark et al, 2011; Meier and Reinagel, 2011), as well as visually-mediated fear conditioning (Shi and Davis, 2001) and eye-reflexes and movements such as nystagmus and saccades (Hess et al.,1985; Hikosaka and Sakamoto, 1987; Fuller, 1985).

The rat and mouse are increasingly important model systems for visual behavior and physiology, it will be important to understand more about the functional organization and connectivity of the early visual pathways in these nocturnal rodents.

**Conclusion**

Our data reveal more anatomical organization in the rat dLGN than previously described. We confirm that inputs from the two eyes are well segregated in the rat dLGN. We find 3-4 geographically-distinct ipsilateral subdomains in the largely-contralateral dLGN. It remains to be determined whether these putative subdomains receive input from distinct classes of retinal ganglion cells or contain duplicate maps of retinotopic space.

**ACKNOWLEDGMENTS**

The authors would like to thank Harvey Karten and Agnieszka Brzozowska-Prechtl of the Karten lab at UCSD in San Diego, CA, for their histological expertise and for the use of lab tools and resources. We would also like to thank Dave Matthews, Jeff Moore, and David Kleinfeld of the Kleinfeld Lab UCSD for generously offering up their advice, technical help, and lab space to the project. We would also like to thank the National Center for Microscopy and Imaging Research (NCMIR) of San Diego, CA, for tools and imaging, as supported by NIH award RR04050 from the National Center for Research Resources. In particluar, Maryann Martone, Hiroyuki Hakozaki, and Stephan Lamont of NCMIR generously provided microscope use and analysis instruction for this project. Finally, we would like to thank Partha Mitra and Vadim Pinskiy of the Mitra lab in Cold Spring Harbor, NY, for their help with data collection.

Funding for this project was provided by National Eye Institute 5R01EY016856 and an award from the James S. McDonnell Foundation. The research was conducted in the absence of any commercial or financial relationships that could be construed as a potential conflict of interest.

**LITERATURE CITED**

Angelucci A., Clascá F., and Sur, M. (1996). Anterograde axonal tracing with the subunit B of cholera toxin: A highly sensitive immunohistochemical protocol for revealing fine axonal morphology in adult and neonatal brains. *J Neurosci Methods* 65(1), 101-112.

Abramoff, M.D., Magalhaes, P.J., and Ram, S.J. (2004). Image Processing with ImageJ. *Biophotonics International* 11(7), 36-42.

Bartlett, E.L. , and Smith, P.H. (1999). Anatomic, Intrinsic, and Synaptic Properties of Dorsal and Ventral Division Neurons in Rat Medial Geniculate Body. *Journal of Neurophys* 81(5), 1999-2016.

Bishop, P.O., Kozak, W., Levick, W.R., and Vakkur, G.J. (1962). The determination of the projection of the visual field on the lateral geniculate nucleus in the cat. *J Physiol Lond* 163, 503-539.

Bowling, D. (1983). Responses to light at different depths in the A layers of the cat's lateral geniculate nucleus. *Invest Opthalmol Suppl* 24, 265.

Brauer, K., Winkelmann, E., Nawka, S., and Strnad, W. (1982). Comparative volumetric investigations on the lateral geniculate body of mammals. *Z Mikrosk Anat Forsch* 96, 400-406.

Braur, K., Schober, W., and Winkelmann, E. (1979). Two morphologically different types of retinal axon terminals in the rats dorsal lateral geniculate nucleus and their relationships to the x and y channel. *Exp Brain Res* 36, 523-532.

Bunt, A.H., Lund, R.D., and Lund, J.S. (1974). Retrograde axonal transport of horseradish peroxidase by ganglion cells of the albino rat retina. *Brain Res* 73, 215-228.

Burn, C. (2008). What is it like to be a rat? Rat sensory perception and its implications for experimental design and rat welfare. *Appl Anim Behav Sci* 112, 1-32.

Carpenter, A.E., Jones, T.R., Lamprecht, M.R., Clarke, C., Kang, I.H., and Friman, O. (2006). CellProfiler: Image analysis software for identifying and quantifying cell phenotypes. Genome *Biol* 7(10), R100.

Carvell, G.E., and Simons, D.J. (1990). Biometric analyses of vibrissal tactile discrimination in the rat. *J Neurosci* 10(8), 2638-2648.

Clark, R.E., Reinagel, P., Broadbent, N.J., Flister, E.D., and Squire, L.R. (2011). Intact performance on feature ambiguous discriminations in rats with lesions of the perirhinal cortex. *Neuron* 70(1), 132-140.

Cleland, B.G., Dubin, M.W., Levick, W.R. (1971). Simultaneous recording of input and output of lateral geniculate neurons. *Nature Lond* 231, 191-192.

Cleland, B.G., Dubin, M.W., and Levick, W.R. (1971). Sustained and transient neurons in the cats retina and lateral geniculate nucleus. *J Physiol Lond* 217, 473-496.

Connolly, M., and Van Essen, D. (1984). The representation of the visual field in parvicellular and magnocellular layers of the lateral geniculate nucleus in the macaque monkey. *Journal of comp Neurol* 226, 544-564.

Discenza, C.B. (2011) "Substructure within the Dorsal Lateral Geniculate Nucleus of the Pigmented Rat". Ph.D. dissertation, University of California, 2011.

Discenza, C.B., Karten, H.J., and Reinagel, P. (2008) Anatomical targets of retinal ganglion cell axons in the Long-Evans rat. Program No. 458.3. 2008 Neuroscience Meeting Planner. Washington, DC: Society for Neuroscience, 2008. Online.

Fukuda, Y. (1977). A three-group classification of rat retinal ganglion cells: Histological and physiological studies. *Brain Res* 119(2), 327-344.

Fuller, J.H. (1985). Eye and head movements in the pigmented rat. *Vision Res* 25(8), 1121-1128.

Gabbott, P.L.A., Somogyi, J., Stewart, M.G., and Hámori, J. (1986). A quantitative investigation of the neuronal composition of the rat dorsal lateral geniculate nucleus using GABA-immunocytochemistry. *Neuroscience* 19(1), 101-111.

Garey, L.J., and Powell, T.P.S. (1968). The projection of the retina in the cat. *J Anat Lond* 102, 189-222.

Grieve, K.L. (2005). Binocular visual responses in cells of the rat dLGN. *J Physiol* 566, 119-124.

Guido, W. (2006). "Cellular Mechanisms Underlying the Remodeling of Retinogeniculate Connections," in *Development and Plasticity in Sensory Thalamus and Cortex*, eds. E. Reha, G. William, and M. Zoltán (New York, NY: Springer), 208-227.

Guillery, R.W. (1970). The laminar distribution of retinal fibers in the dorsal lateral geniculate nucleus of the cat. *J Comp Neurol* 138, 339-368.

Hamos, J.E., Van Horn, S.C., Raczkowski, D., Uhlrich, D.J., and Sherman, S.M. (1985). Synaptic connectivity of a local circuit neurone in lateral geniculate nucleus of the cat. *Nature* 317, 618-621.

Hayhow, W.R., Sefton, A., and Webb, C. (1963). Primary optic centers of the rat in relation to the terminal distribution of the crossed and uncrossed optic nerve fibers. *J Comp Neurol* 118(3), 295-321.

Hess, B,J., Precht, W., Reber, A., and Cazin, L. (1985). Horizontal optokinetic ocular nystagmus

in the pigmented rat. *Neuroscience* 15(1), 97-107.

Hickey, T.L., and Spear, P.D. (1976). Retinogeniculate projections in hooded and albino rats: An autoradiographic study. *Exp Brain Res* 24(5), 523-529.

Hikosaka, O., and Sakamoto, M. (1987). Dynamic characteristics of saccadic eye movements in the albino rat. *J Neurosci Res* 4(4), 304-308.

Hill, A.J., and Best, P.J. (1981). Effects of deafness and blindness on the spatial correlates of hippocampal unit activity in the rat. *Exp Neurol* 74(1), 204-217.

Holscher, C., Schnee, A., Dahmen, H., Setia, L., and Mallot, H.A. (2005). Rats are able to navigate in virtual environments. *J Exp Biol* 208, 561-569.

Huberman, A.D., Manu, M., Koch, S.M., Susman, M.W., Lutz, A.B., and Ullian, E.M. (2008). Architecture and activity-mediated refinement of axonal projections from a mosaic of genetically identified retinal ganglion cells. *Neuron* 59(3), 425-438.

Huberman, A.D., Wei, W., Elstrott, J., Stafford, B.K., Feller, M.B., and Barres, B.A. (2009). Genetic identification of an on-off direction- selective retinal ganglion cell subtype reveals a layer-specific subcortical map of posterior motion. *Neuron* 62(3), 327-334.

Hutson, K.A., and Masterton, R.B. (1986). The sensory contribution of a single vibrissa's cortical barrel. *J Neurophysiol* 56(4), 1196-1223.

Jacobs, G.H., Fenwick, J.A., and Williams, G.A. (2001). Cone-based vision of rats for ultraviolet and visible lights. *J Exp Biol* 204, 2439-2446.

Jones, E.G. (2007). *The Thalamus*. New York: Cambridge University Press.

Kaas, J.H., Guillery, R.W., and Allman, J.M. (1972). Some principles of organization in the dorsal lateral geniculate nucleus. *Brain Behav Evol* 6, 253-299.

Kinston, W.J., Vadas, M.A., and Bishop, P.O. (1969). Multiple projection of the visual field to the medial portion of the dorsal lateral geniculate nucleus and the adjacent nuclei of the thalamus of the cat. *J comp neur* 136, 295-316.

Kulvicius, T., Tamosiunaite, M., Ainge, J., Dudchenko, P., and Worgotter, F. (2008). Odor supported place cell model and goal navigation in rodents. *J Comput Neurosci* 25(3), 481-500.

Laties, A.M., and Sprague, J.M. (1966). The projection of optic fibers to the visual centers in the cat. *J Comp neurol* 137, 35-70.

Levick, W.R., Cleland, B.G., and Dublin, M.W. (1972). Lateral geniculate neurons of cat: retinal inputs and physiology. *Invest Opthalmol* 11(5), 302-311.

Lund, R.D., and Cunningham, T.J. (1972). Aspects of synaptic and laminar organization of the mammalian lateral geniculate body. *Invest Ophthalmol* 11(5), 291-302.

Maaswinkel, H., and Whishaw, I.Q. (1999). Homing with locale, taxon, and dead reckoning strategies by foraging rats: Sensory hierarchy in spatial navigation. *Beh Brain Res* 99(2), 143-152.

Major, D.E., Rodman, H.R., Libedinsky, C., and Karten H.J.(2003) Pattern of retinal projections in the California ground squirrel (Spermophilus beecheyi): anterograde tracing study using cholera toxin. J Comp Neurol.463(3):317-40.

Malpeli, J.G., and Baker, F.H. (1975). The representation of the visual field in the lateral geniculate nucleus of Macaca Mulatta. *J Comp Neurol* 161, 569-594.

Marc, R.E., and Jones, B.W. (2002). Molecular phenotyping of retinal ganglion cells. *J Neurosci* 22(2), 413-427.

Martin, P.R. (1986). The projection of different retinal ganglion cell classes to the dorsal lateral geniculate nucleus in the hooded rat. *Exp Brain Res* 62(1), 77-88.

Matteau, I., Boire, D., and Ptito, M. (2003). Retinal projections in the cat: A cholera toxin B subunit study. *Vis Neurosci* 20(5), 481-493.

Meier, P., Flister, E.D. and Reinagel, P. (2011). Collinear features impair visual detection by rats. *J Vis* 11(3), 1-16.

Meier, P., and Reinagel, P. (2011). Rat performance on visual detection task modeled with divisive normalization and adaptive decision thresholds. *J Vis* 11(9), 1-17.

Montero, V.M., Brugge, J.F., and Beitel, R.E. (1968). Relation of the visual field to the lateral geniculate body of the albino rat. *J Neurophys* 31(2), 221-236.

Morris, R. (1984). Developments of a water-maze procedure for studying spatial learning in the rat. *J Neurosci Methods* 11(1), 47-60.

Murray, K.D., Rubin, C.M., Jones, E.J., and Chalupa, L.M. (2008). Molecular correlates of laminar differences in the Macaque dorsal lateral geniculate nucleus. *J Neurosci* 28(46), 12010-12022.

Najdzion, J., Wasilewska, B., Bogus-Nowakowska, K., Równiak, M., Szteyn, S., and Robak, A. (2009). A morphometric comparative study of the lateral geniculate body in selected placental mammals: the common shrew, the bank vole, the rabbit, and the fox. *Via Medica* 68(2), 70-78.

Paxinos, G., and Watson, C. (1998). *The Rat Brain in Stereotaxic Coordinates, 4th ed.* San Diego, CA: Academic Press.

Prusky, G.T., Harker, K.T., Douglas, R.M., and Whishaw, I.Q. (2002). Variation in visual acuity within pigmented, and between pigmented and albino rat strains. *Beh Brain Res* 136(2), 339-348.

Reese, B.E. (1984). The projection from the superior colliculus to the dorsal lateral geniculate nucleus in the rat. *Brain Res* 305(1), 162-168.

Reese, B.E. (1988). 'Hidden lamination' in the dorsal lateral geniculate nucleus: The functional organization of this thalamic region in the rat. *Brain Res Rev* 13(2), 119-137.

Reese, B.E., and Cowey, A. (1983). Projection lines and the ipsilateral retino-geniculate pathway in the hooded rat. *Neuroscience* 10(4), 1233-1247.

Reese, B.E., and Jeffery, G. (1983). Crossed and uncrossed visual topography in dorsal lateral geniculate nucleus of the pigmented rat. *J Neurophys* 49(4), 877-885.

Reiner, A., Zhang, D., and Eldred, W.D. (1996). Use of the sensitive anterograde tracer cholera toxin fragment B reveals new details of the central retinal projections in turtles. *Brain Behav Evol* 48(6), 307-337.

Roe, A.W., Garraghty, P.E., and Sur, M. (1989). Terminal arbors of single on center and off center x and y retinal ganglion cell axons within the ferrets lateral geniculate nucleus. *J Comp Neurol* 288, 208-242.

Roy, S., Jayakumar, J., Martin, P.R., Dreher, B., Saalmann, Y.B., Hu, D., and Vidyasagar, T.R. (2009). Segregation of short-wavelength-sensitive (S) cone signals in the macaque dorsal lateral geniculate nucleus. *Eur J Neurosci* 30(8), 1517-1526.

Sackett, G.P., Graham, J., and DeVito, J.L. (1989). Volumetric growth of the major brain divisions in fetal macaca nemestrina. *J Hirnforsch* 30(4), 479-87.

Sanderson, K.J. (1971). Visual field projection columns and magnification factors in the lateral geniculate nucleus of the cat. *Exp Brain Res* 13, 159-177.

Sanderson, K.J. (1971). The projection of the visual field to the lateral geniculate and medial interlaminar nuclei in the cat. *J Comp Neurol* 143(1), 101-117.

Save, E., Nerad, L., and Poucet, B. (2000). Contribution of multiple sensory information to place field stability in hippocampal place cells. *Hippocampus* 10(1), 64-76.

Schiller, P.H., and Malpeli, J.G. (1978). Functional specificity of lateral geniculate nucleus laminae of the rhesus monkey. *J neurophysiol* 41, 788-797.

Shapley, R.V., and Perry, H. (1986). Cat and monkey retinal ganglion cells and their visual functional roles. *Trends Neurosci* 9, 229-235.

Sherman, S.M., and Guillery, R.W. (2000). *Exploring the Thalamus*. San Diego, CA: Academic Press.

Sherman, S.M., and Spear, P.D. (1982). Organization of visual pathways in normal and visually deprived cats. *Physiol Rev* 62(2), 738-855.

Sherman, S.M., and Guillery, R.W. (1996). Functional organization of thalamocortical relays. *J Neurophys* 76(3), 1367-1395.

Shi, C., and Davis, M. (2001). Visual pathways involved in fear conditioning measured with fear-potentiated startle: Behavioral and anatomic studies. *J Neurosci* 21(24), 9844-9855.

So, K.F., Campbell, G., and Lieberman, A.R. (1990). Development of the mammalian retinogeniculate pathway: target finding, transient synapses, and binocular segregation. *J Exp Biol* 153, 85-104.

Stryker, M.P., and Zahs, K.R. (1983). On and off sublaminae in the lateral geniculate nucleus of the ferret. *J Neurosci* 3, 1943-1951.

Szmajda, B.A., Buzas, P., PitzGibbon, T., and Martin, P.R. (2006). Geniculocortical relay of blue off signals in the primate visual system. *Proc Natl Acad Sci* 103, 19512-19517.

Thompson, W.R., and Solomon, L.M. (1954). Spontaneous pattern discrimination in the rat. *J Comp Physiol Psychol* 47(2), 104-107.

Torborg, C.L., and Feller, M.B. (2004). Unbiased analysis of bulk axonal segregation patterns. *J Neurosci Methods* 135(1-2), 17-26.

Villena, A., Diaz, F., Requena, V., Chavarria, I., Rius, F., and Perez de Vargas, I. (1997). Quantitative morphological changes in neurons from the dorsal lateral geniculate nucleus of young and old rats. *Anat Rec* 248(1):137-141.

Wu, C., Russel, R.M., and Karten, H.J. (1999). The transport rate of cholera toxin B subunit in the retinofulgal pathway of the chick. J Neurosci 92(2): 665-676.

Yonehara, K., Ishikane, H., Sakuta, H., Shintani, T., Nakamura-Yonehara, K., Kamiji, N.L., Usui, S., and Noda, M. (2009). Identification of retinal ganglion cells and their projections involved in central transmission of information about upward and downward image motion. *PloS One* 4(1), e4320.

Zoccolan, D., Oertelt, N., DiCarlo, J.J., Cox, D.D. (2009). A rodent model for the study of invariant visual object recognition. *Proc Natl Acad Sci USA* 106(21), 8748-8753.